\documentclass[12pt,preprint,fleqn]{aastex}
\usepackage{epsfig}
\usepackage{amsmath}
\usepackage{mathptmx}
\def\hmpc		{{~$h^{-1}$~Mpc}}
\def\lcdm               {{$\Lambda$CDM}}
\begin{document}
\title{\Large{Crawling the Cosmic Network: Exploring the Morphology of Structure in the Galaxy Distribution}}
\author{Nicholas A. Bond\altaffilmark{1}, Michael A. Strauss, Renyue Cen}
\affil{Princeton University}
\affil{Princeton University Observatory, Princeton, NJ 08544\label{Princeton}}
\altaffiltext{1}{nbond@physics.rutgers.edu}
\clearpage
\begin{abstract} 
Although coherent large-scale structures
such as filaments and walls are apparent to the eye in galaxy redshift
surveys, they have so far proven difficult to characterize with
computer algorithms.  This paper presents a procedure that uses the
eigenvalues and eigenvectors of the Hessian matrix of the galaxy
density field to characterize the morphology of large-scale
structure.  By analysing the smoothed density field and its Hessian
matrix, we can determine the types of structure -- walls, filaments,
or clumps -- that dominate the large-scale distribution of galaxies
as a function of scale.  We have run the algorithm on mock galaxy
distributions in a \lcdm~cosmological N-body simulation and the
observed galaxy distributions in the Sloan Digital Sky Survey.  The
morphology of structure is similar between the two catalogues, both
being filament-dominated on $10$--$20$\hmpc~smoothing scales and
clump-dominated on $5$\hmpc~scales.  There is evidence for walls in
both distributions, but walls are not the dominant structures on
scales smaller than $\sim 25$~$h^{-1}$~Mpc.  Analysis of the
simulation suggests that, on a given comoving smoothing scale,
structures evolve with time from walls to filaments to clumps, where
those found on smaller smoothing scales are further in this
progression at a given time.  
\end{abstract}
\onecolumn
\section{Introduction}
Inflationary models of the early universe
\citep{Inflation} suggest that the large-scale distribution of
matter should be well described by a gaussian random field, the
statistical properties of which can be completely specified by the
power spectrum.  As structure evolves, however, non-linear effects
become increasingly important and higher-order statistics must be
computed if we wish our description to be complete.  A number of
studies have been performed to measure the three-point correlation
function of galaxies
\citep[e.g.][]{Peebles3PCF,3PCF2,3PCFSDSS,3PCF2dF} and its
Fourier Transform (the bispectrum), as well as $N$-point statistics
\citep[e.g.,][]{NPCF,Verde00,NPCFSDSS}.  However, the filaments, walls, and
clusters that appear in the non-linear regime produce very
high-order correlations which are difficult to compute and
interpret.

Three-dimensional maps of the galaxy distribution show walls and
filaments
\citep[e.g.,][]{Thompson78,CfaSurvey,GreatWall,2dF,SloanGreatWall}, as
do cosmological N-Body simulations
\citep[e.g.,][]{Davis85,CosmicWeb,MMF,Hahn07}.  At present, there are a number of
methods for quantifying filamentarity
\citep[e.g.,][]{Percolation,Shandarin98,Starck04,Shen06,Skeleton},
as well as for tracing individual filaments \citep[e.g.,][]{Skeleton2,Spine,SCP09,Thesis2}.
Filaments produce deep potential wells on large scales, surpassed only
by clumps\footnote[1]{Clumps are defined as nearly spherical
overdensities that have undergone significant collapse along all three
principal axes.  Bound clusters are a subset of this class of
structure.}.  As such, they will present a signal for gravitational
lensing studies on the largest scales \citep{LensingFils2,
LensingFils}.  Furthermore, filaments can act as a pathway for matter
accreting onto young galaxies and galaxy clusters
\citep[e.g.][]{CL0016, FilGalForm} and can align the spins of dark
matter haloes \citep{Hahn07}.

In the classic picture put forward by \citet{Pancake}, an ellipsoidal
overdensity will collapse along its shortest axis first, forming a
`pancake'-like structure.  Some early models of structure
formation, such as the `hot dark matter' models, led to a
`top-down' scenario, in which an excess of power at large scales
led to a universe filled with giant sheets, extending for tens to
hundreds of megaparsecs \citep{Zelly82, Frenk83}.  We now think that
dark matter is cold and that structure formed in a bottom-up fashion,
with the smaller structures forming before the larger ones.  Walls may
still exist, but they are not as obvious in either the galaxy catalogues
or cosmological simulations as they would be in the hot dark matter
models. Furthermore, it is not clear that we can apply the simple
picture of ellipsoidal collapse to structures that are experiencing
tidal forces from nearby structures and continual fragmentation on
smaller scales. 

One approach to describing bottom-up structural evolution is known as
the `peak--patch' theory \citep{PeakPatch}, in which clumps, the
rare overdensities in a smoothed density field, form first.
Filament-like features then arise along the longest principal axis of
the tidal tensor, forming bridges between the clumps, followed by
wall-like structures connecting the bridges.  Although this
description incorporates tidal interactions between neighboring
structures, it focuses on only one smoothing scale at a time and is
forced to resort to increasingly complex $N$-point statistics to
describe the development of structures on scales much larger than the
smoothing length.

In this paper, we are focused on measuring the dimensionality of structures in
the overall distribution of matter as a function of scale.  Past
studies of the correlation function have shown large-scale structure
to be consistent with a fractal on $\lesssim 10$\hmpc~scales
\citep{Fractal1,Fractal2}, suggesting that many of the structures
apparent to the eye are nested into larger structures.
Any measure of structure that ignores this fact is giving an
incomplete picture of the matter distribution. 

In what follows, we will describe a procedure to quantify the
prominence of structures, such as filaments or walls, in the
large-scale distribution of galaxies and then apply this procedure
both to simulations and to observations of the large-scale
distribution of galaxies from the Sloan Digital Sky Survey
\citep[SDSS;][]{SDSS}.  First, in \S~\ref{sec:Method}, we will define
the local structural parameters.  Walls, filaments, and clumps will each
have a unique fingerprint in the resulting parameter space
($\lambda$-space) and we will present these fingerprints, along with
their dependence on the properties of a given structure, in
\S~\ref{sec:ParDist}.  In \S~\ref{sec:Sims}, we describe the N-body
simulations and mock galaxy catalogues with which we will develop our
statistics and compare the large-scale structure data.  In
Sections~\ref{sec:GRFs} and \ref{sec:DSpace}, we discuss the
$\lambda$-space distributions of gaussian random fields and dark
matter in a cosmological simulation, respectively.  We will examine
these statistics and compare these distributions to those of real
galaxy data in \S~\ref{sec:Data}.  Finally, in
\S~\ref{sec:Discussion}, we will summarize the results and discuss
their implications for our understanding of large-scale structure.
Bond, Strauss, \& Cen (Paper II, in preparation) will describe a method of finding individual
filamentary structures.

\section{Method}
\label{sec:Method}
\subsection{  Single-scale smoothing of the density field and its Hessian  }
\label{subsec:denshess}

Filaments, clusters, and walls all present sharp features in the
density field along at least one of their principal axes.  If we wish
to bring out such features, we should generate not only the density
field, but also its second derivatives.  The smoothed density field is defined
as,
\begin{equation} 
\label{eq:Density} 
\tilde{\rho}(\boldsymbol{x})=\int f(\boldsymbol{x}-\boldsymbol{x'})\rho(\boldsymbol{x'}){\rm d}^3\boldsymbol{x'}, 
\end{equation}
where $\rho(\boldsymbol{x})$ is the density field and $f(\boldsymbol{x})$ is
the smoothing kernel.  The density field is initially composed of a
sum of delta functions (at the positions of the galaxies or dark
matter particles), which are zero-valued at infinity, so integration
by parts yields the smoothed Hessian (i.e. matrix of second partial
derivatives),
\begin{equation} 
\label{eq:SmoothHessian} 
\tilde{H}_{{\rm ij}}(\boldsymbol{x})=\int f(\boldsymbol{x}-\boldsymbol{x'})\frac{{\rm \partial}^2 \rho(\boldsymbol{x'})}{{\rm \partial} x_{\rm i}'{\rm \partial} x_{\rm j}'}{\rm d}^3\boldsymbol{x'}=-\int\frac{{\rm \partial}^2 f(\boldsymbol{x}-\boldsymbol{x'})}{{\rm \partial} x_{\rm i}'{\rm \partial} x_{\rm j}'}\rho(\boldsymbol{x'}){\rm d}^3\boldsymbol{x'}.
\end{equation}
A gaussian smoothing kernel will be used throughout this paper.

The convolution is performed on a discrete grid after first performing
a Fast Fourier Transform (order $N{\rm log} N$) on the kernel and data
vectors.  This operation implicitly assumes periodic boundary
conditions, a criterion that is met by the simulation data (which is
in a periodic box), but not by the survey data.  For the latter case,
we will only search for structures more than two smoothing lengths
from the box edge.

The idea to use the Hessian matrix to characterize local large-scale structure was pioneered by \citet{Colombi01} and later used for the identification of individual large-scale structures by \citet{MMF}.  A similar method, which uses the eigenvalues of the tidal shear field, was implemented by \citet{Hahn07} to characterize large-scale structure in the vicinity of dark matter halos in N-body simulations. This technique was later applied to galaxy redshift surveys by \citet{LL08}.  The method presented here is also similar to that used by \citet{Forero09} to present a dynamical classification of large-scale structures.  

It has been suggested by some authors \citep[e.g.][]{AdaptSmooth} that, because filaments present themselves on a multitude of scales, one should {\it adaptively} smooth the galaxy density field.  It's true that a visual inspection of a high-resolution density map of large-scale structure will reveal filaments on scales as small as $\sim 1$~\hmpc, but these filaments will often be {\it embedded} in larger structures, which will be difficult to identify in a single density map that deconstructs them into their smaller components (see \S~\ref{subsubsec:Scale} for further discussion).  For these reasons, we will smooth {\it separately} on a series of length scales when searching for filaments and walls.

\subsection{  The $\lambda$ parameters and axis of structure }
\label{subsec:shmaffpars}

The Hessian matrix can be visualized as an ellipsoid with the major
axis aligned along the direction of lowest concavity.  In filamentary
regions, this direction will be {\it along} the filament
itself, so the local orientation of a filament is given by the major
axis of the Hessian ellipsoid at that point.  By diagonalizing the
Hessian matrix, we move into the coordinate system of this ellipsoid
and obtain its principal axes (the Hessian eigenvalues) and
orientation (eigenvectors).  

The Hessian eigenvalues at a given grid cell will be denoted
$\lambda_{\rm i}$ and ordered such that $\lambda_1<\lambda_2<\lambda_3$.
Along with the density and gradient vector, these provide an
indication of the shape and contrast of local structure.  The
orientation of the structure at a given grid cell is given by
$\boldsymbol{A}_3$, the eigenvector corresponding to $\lambda_3$.
Although real filaments will often have a non-negligible gradient
along their length, we will consider only the {\it axis} along the
filament, hereafter referred to as the `axis of structure'.  We wish
to keep our notion of a filament very simple -- locally, it is a
structure that is concave down along two principal axes and nearly
flat along the other one.  We will make no assumptions about the sign
or magnitude of the first derivative along its length, but for the
sake of convention, we will define $\boldsymbol{A}_3$ such that
\begin{equation}
\label{eq:CurveDir} 
\boldsymbol{A}_3 \mathbf{\cdot} \boldsymbol{\nabla}\rho>0.
\end{equation} 

We define the dimensionless eigenvalues,
\begin{equation}
\lambda_{\rm i}'\equiv\frac{l^2\lambda_{\rm i}}{\bar{\rho}},
\end{equation}
where $l$ is the smoothing length and $\bar{\rho}$ is the mean
density.  Near the centres of filaments, where the density field is
concave downalong two axes of the filament, we expect the following
relationships between the $\lambda'$-parameters:
\begin{align}
\label{filcrit3d}
&\lambda_1'\ll 0 \nonumber \\
&\lambda_2'\sim \lambda_1' \ll 0 \\
&|\lambda_3'|\ll |\lambda_1'|. \nonumber
\end{align}
For walls, the density field is concave down along only one axis and
slowly varying along the other two,
\begin{align}
\label{wallcrit3d}
&\lambda_1'\ll 0 \nonumber \\
&|\lambda_2'|\ll |\lambda_1'| \\
&|\lambda_3'|\ll |\lambda_1'|. \nonumber
\end{align}

Filaments are essentially
one-dimensional objects, so a single vector can be used to represent
their orientation.  However, if we wish to consider walls, an additional
eigenvector, $\boldsymbol{A}_2$, is needed.  The full set of parameters 
to describe the geometry of structure at a given cell is given by
\begin{equation}
\label{shmaffset3}
\boldsymbol{S}=(\boldsymbol{A}_2,\boldsymbol{A}_3,\lambda_1',\lambda_2',\lambda_3').
\end{equation}

\section{Geometry with $\lambda$-space distributions}
\label{sec:ParDist}

In Equations {\ref{filcrit3d}} and
{\ref{wallcrit3d}} are crude relations that we expect the $\lambda'$
parameters to satisfy in the vicinity of specific large-scale structures.
This can be made a bit more quantitative -- filaments and walls
will have simple `fingerprints' in
$\lambda_1'$-$\lambda_2'$-$\lambda_3'$ space (hereafter,
$\lambda$-space) that we can look for in real data.

\subsection{Three-dimensional fingerprints}
\label{subsubsec:3DFinger}

\begin{figure}[t]
\plotone{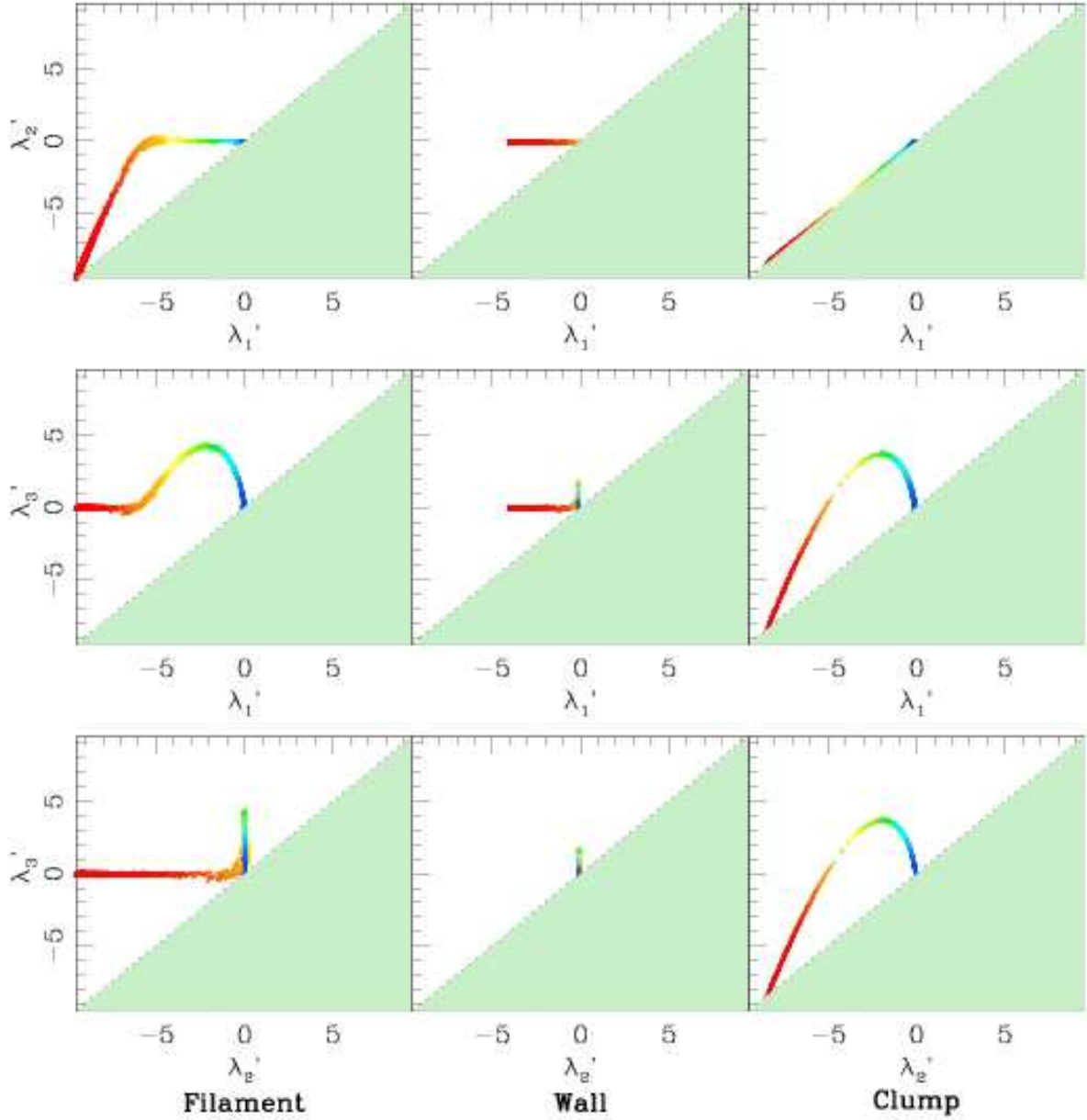}
\caption{The structural `fingerprints' for, from right-to-left, an
isolated filament, a wall, and a cluster in an otherwise uniform
background, displayed in three projections of $\lambda$-space.  The
points are colour-coded by local density (blue, green, yellow, orange,
and red in order of increasing density) and inaccessible regions are
shaded light gray, with eigenvalues defined such that
$\lambda_1'<\lambda_2'<\lambda_3'$.  All structures are smoothed on a
scale much larger than their width.\label{fig:OneObjBox3}}
\end{figure}

In Fig.~\ref{fig:OneObjBox3}, we plot the $\lambda$-space
projections for point distributions drawn from a two-dimensional
(left), one-dimensional (centre), and three-dimensional gaussian
function (right).  These are models of a filament, wall, and clump,
respectively, and are each placed in a $200$\hmpc~box.  Smoothing will
be done on a scale larger than the width of the structure, so the mock
structures approximate a line, a plane, and a point.  We confirmed
that the fingerprints are coordinate-independent by rotating the mock
structures to random orientations.

Fig.~\ref{fig:OneObjBox3} plots the distribution of the $\lambda$
parameters evaluated at the position of each galaxy, each interpolated
on a $128 \times 128 \times 128$ grid.  The points in this figure are
colour-coded by local density (blue, green, yellow, orange, and red in
order of increasing density).  Note that the definition, $\lambda_1'\leq
\lambda_2'\leq \lambda_3'$, entirely excludes four octants in
$\lambda$-space, as well as portions of the remaining octants
(delimited by the dashed lines in Fig.~\ref{fig:OneObjBox3}).  In
this and all subsequent multidimensional interpolations, we use a
third-order polynomial routine given in \citet{NumericalRecipe}.

In all of the $\lambda$-space distributions, there is a tight
grouping of blue points near the origin made up of background objects
well away from the mock structures; the Hessian matrix of a uniform
density field is all zeros.  The remaining points trace the
`fingerprint' of the wall, filament, or clump.  Each fingerprint can
be understood by considering the curvature properties of a gaussian
function.  A gaussian function has curvature given by
\begin{equation}
\label{eq:gausscurve}
\frac{{\rm \partial}^2 f}{{\rm \partial} x^2}=f(x)\left(\frac{x^2}{\sigma^4}-\frac{1}{\sigma^2}\right),
\end{equation}
where $f(x)$ is a one-dimensional gaussian function with width,
$\sigma$.  At $x=0$, the right-hand side of
Equation~\ref{eq:gausscurve} is equal to $-\rho(x)/\sigma^2$.  Moving
to larger values of $|x|$, the value of this function increases,
reaching a maximum at $|x|=\sqrt{3}\sigma$, and then switching sign at
$|x|=\sigma$.  Beyond this, the curvature of a gaussian asymptotically
approaches zero.  The fingerprint of the wall follows naturally from
this simple analysis.  At all points near the wall centre (the densest
regions, in red), the axis of structure will be either aligned with or
orthogonal to the x coordinate axis -- the density field has no y
or z dependence so the curvature of the density field along those
axes will always be zero.  This means that $\frac{{\rm \partial}^2
\rho}{{\rm \partial} x^2}$ will correspond to either $\lambda_1'$ (the
smallest eigenvalue) or $\lambda_3'$ (the largest eigenvalue),
depending upon the sign of the curvature along the x direction
(which follows Equation~\ref{eq:gausscurve} after smoothing).
Therefore, for $|x|<\sigma$
\begin{align}
\label{xltsig}
&\lambda_1=\frac{{\rm \partial}^2 \rho}{{\rm \partial} x^2}<0 \\
&\frac{{\rm \partial}^2 \rho}{{\rm \partial} y^2}=\frac{{\rm \partial}^2 \rho}{{\rm \partial} z^2}=\lambda_2=\lambda_3=0  \nonumber
\end{align}
and for $|x|>\sigma$,
\begin{align}
\label{xgtsig}
&\frac{{\rm \partial}^2 \rho}{{\rm \partial} y^2}=\frac{{\rm \partial}^2 \rho}{{\rm \partial} z^2}=\lambda_1=\lambda_2=0 \\
&\lambda_3=\frac{{\rm \partial}^2 \rho}{{\rm \partial} x^2}>0.  \nonumber
\end{align}
This explains the origin of the half-cross shape in the centre panel
of Fig.~\ref{fig:OneObjBox3}. 

For the clump, the direction of least curvature (and, therefore, the
axis of structure) will be either parallel or perpendicular to the
radial vector from the clump centre.  Along this radial vector, the
density field will follow the curvature properties of a gaussian
function, corresponding to $\lambda_3'$ in the bottom-right panel of
Fig.~\ref{fig:OneObjBox3}.  Orthogonal to this vector, the density
field will always have negative curvature, meaning that the smallest
two eigenvalues ($\lambda_1'$ and $\lambda_2'$, see the top-right
panel) will always be negative and, by symmetry, equal in magnitude.
All three eigenvalues increase with distance from the clump centre (in
red) until the curvature of the gaussian function maximizes at
$r=\sqrt{3}\sigma$.  Beyond this, $\lambda_3'$ drops and the
fingerprint approaches the origin.

Finally, in the mock filament (left column of
Fig.~\ref{fig:OneObjBox3}), there is no z-dependence in the
density field , so one of the $\lambda$-values will always be zero
and the fingerprint will be restricted to one of the $\lambda$-space
coordinate planes.  Near the filament centre (in red), the curvature
is negative orthogonal to the filament axis.  This means that the two
non-zero eigenvalues are negative and the fingerprint lies in the
$\lambda_2'$--$\lambda_1'$ plane.  Both of these eigenvalues will
increase with distance from the filament centre, with the one
corresponding to the radial eigenvector following the curvature of the
one-dimensional gaussian function (Equation~\ref{eq:gausscurve}).  At
a cylindrical radius, $R=\sqrt{3}\sigma$, the curvature changes sign
and this eigenvalue becomes positive (and, therefore, becomes
$\lambda_3'$), after which the fingerprint remains on the
$\lambda_3'$--$\lambda_1'$ plane and behaves like a projection of the
clump fingerprint.

\subsection{Wall and filament proportions}
\label{subsubsec:Proportions}

Neither the real data nor the cosmological simulations will have such
a simple structure as the idealized models described the last section.
To improve the realism of the tests, we will give the walls and
filaments a finite extent, as well as place multiple structures within
a single box.

The first set of tests \citep[see ][]{Thesis} varied the length of the
filament, ranging from six times to twice the smoothing length.  As
the filament size was reduced, the filament began to resemble a clump
in $\lambda$-space.  Filaments also have a wide range in width, as
the very largest structures in the universe \citep[e.g., the Sloan
Great Wall,][]{SloanGreatWall} have thicknesses of $\sim 50$\hmpc~in
redshift space, much greater than the widths of `typical' filaments
visible in redshift surveys. We generated sample filaments to test
this width dependence, varying it between half and twice the smoothing
length.  We found that, even for very long filaments, the
$\lambda$-space distributions show few of the characteristic
`filament' features unless the structure is narrower than the
smoothing length.  Thus, single-scale smoothing is insensitive to
structures much wider than the smoothing length, suggesting that it
will be useful to measure the {\it scale-dependence} of filamentarity
with a series of measurements on different smoothing scales.


We have also considered the impact of nearby structures on the filament fingerprint.  We find that fingerprint remains largely unchanged so long as the density of filaments is not so high that nearby structures are within one smoothing length of one another.  In this limit, a smoothing length less than the average separation between structures is needed to distinguish them.

We conducted similar tests for walls, varying their side lengths,
widths, and number within a box.  Points in the vicinity of a wall's
edge will have $\lambda$-parameters similar to those of a filament
and parameters for points near the corners will be like those of the
clump.  As with the filaments, a great deal of scatter was introduced
when the structures were made very wide or very numerous.

Although the morphology of the $\lambda$-space distributions can vary
with a structure's width, length, or overdensity, there are
discriminating features for each type of structure.  Filaments can
most easily be distinguished from clumps by a concentration
around the $\lambda_3'=0$ axis in the $\lambda_3'$--$\lambda_1'$
projection, while walls can be distinguished from filaments by a
concentration on the $\lambda_2'=0$ axis in the
$\lambda_2'$--$\lambda_1'$ projection.  In \S~\ref{sec:Data}, these
facts will be used to probe structure in cosmological simulations and
SDSS galaxy data.

\section{ The simulation }
\label{sec:Sims}

Before we present the $\lambda$-space projections of the observed
galaxy distribution, it is useful to explore the predictions of the
standard cosmological model.  A cosmological simulation allows us to
develop our techniques on a data set not subject to redshift
distortions, sparse sampling, and complicated window functions.
Furthermore, we can simulate these effects in mock catalogues, providing
a useful standard for comparison to the real SDSS data.

We use a series of cosmological CDM $N$-body simulations with
$\Omega_{\rm m} = 0.29$, $\Omega_\Lambda = 0.71$, $\sigma_8 = 0.85$, and $h
= H_0/(100 $~km~s$^{-1}$~Mpc$^{-1}$)$=0.69$ \citep{WMAP2}.  The
simulation is performed within a
$200$\hmpc~box with $512^3$ particles, each with mass, $m_{\rm p} = 4.77
\times 10^9$~$h^{-1}$~M$_{\rm \sun}$, and is run on a $92$-node Beowulf
cluster located at Princeton University.  The positions and velocities
of dark matter particles are output at six redshifts -- $z=0$, $0.3$,
$0.5$, $1$, $2$, and $3$.  Because of memory restrictions, we split
the simulation box into a $3 \times 3 \times 3$ lattice in order to
run a halo-finder and identify galaxies.  Each subbox was padded with
the particle data within $10$\hmpc~of its boundaries.  The padding
size is an order of magnitude larger than the largest dark matter
haloes, so edge effects are negligible.

We identified
dark matter haloes with the \small{HOP} algorithm \citep[hereafter
EH98]{HOP}, which searches for peaks in the density field by moving
from particle to particle until it reaches one that is in a denser environment than any
of its neighbors.  Densities are computed with an adaptive kernel with
a length scale set by the $N_{\rm dens}$ nearest particles, and the
`hopping' can occur to any of the $N_{\rm hop}$ nearest particles.
Groups are then pruned, merged or disbanded depending on their peak
and outer density thresholds (see EH98 for more details).  Following
EH98, we select $N_{\rm hop}=16$, $N_{\rm dens}=64$, $N_{\rm merge}=4$,
$\delta_{\rm peak}=240$, $\delta_{\rm saddle}=200$, and $\delta_{\rm outer}=80$.
In addition, we keep only haloes with $N \ge 8$ particles.  These
parameters give halo distributions similar to those of a
friends-of-friends algorithm \citep{FOF} with a linking length of $0.2$.  We note
that our results will be insensitive to the choice of the minimum halo mass
as the faintest galaxies we will be studying (see \S~\ref{subsubsec:GalSimDspace})
have a mean halo mass of $\sim 10^{12}$~M$_{\sun}$, corresponding to $N \simeq 200$ particles.

\section{ Gaussian random fields}
\label{sec:GRFs}

In both the cosmological simulations and observed galaxy distribution, we wish to identify the morphological features that are unique to non-linear structure evolution; that is, features which are unlikely to appear in a gaussian random field.  If we wish to reveal these features in $\lambda$-space, then we should compare the Hessian parameters in observed or simulated data to those in a gaussian random field with the same power spectrum.  For the non-linear power spectrum, we use the prescription of \citet{NonLinearPS}, with the concordance model parameters of \citet{WMAP2}.

\begin{figure}[t]
\epsscale{1.0}
\plotone{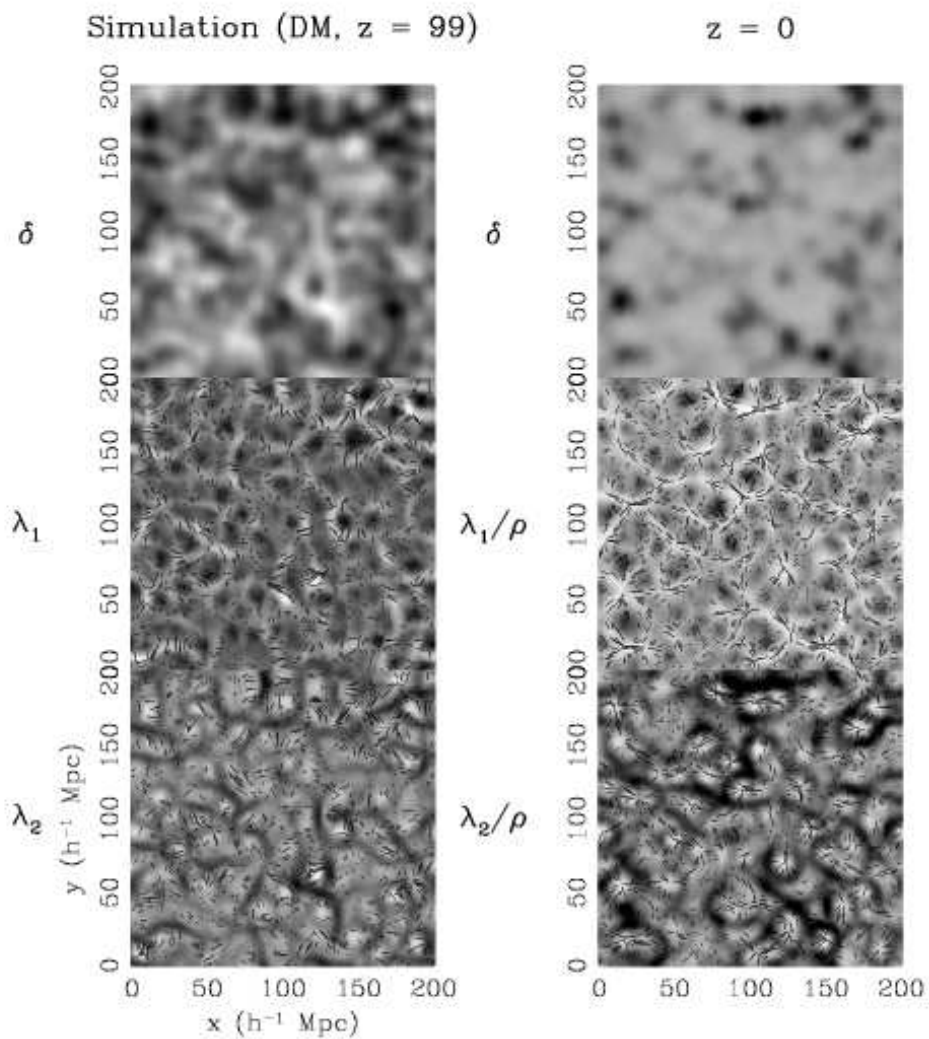}
\caption{
A slice ($10$\hmpc~deep) from the dark matter initial conditions of a cosmological simulation at $z=99$ (left column) compared with the same slice at $z=0$ (right column).  The top row shows the density
fields smoothed on a $5$\hmpc~scale, while the middle and bottom rows
show the corresponding $\lambda_1'$ and $\lambda_2'$ maps, respectively.
The $\lambda$ maps have been normalized to the local density for the $z=0$ dark matter distribution in order to emphasize the large-scale distribution of structures and provide an easy comparison to the initial conditions in the left column.  Bars have been plotted over the $\lambda$-maps to indicate the
direction of the axis of structure at random points on the plane.  The
bar's length is proportional to the magnitude of $\lambda_1'$ at that
point.  It is clear to the eye that the axis of structure aligns with the local filamentary structure in the dark matter distributions.  This alignment is absent in the initial conditions, despite a similar $\lambda_1'$ distribution.
\label{fig:GRFCompare2d_6}}
\end{figure}

The left column of Fig.~\ref{fig:GRFCompare2d_6} shows a slice from the $z=99$ initial conditions for the dark matter distribution in a cosmological simulation, smoothed on a $5$~\hmpc~scale.  In the top panel we plot the smoothed density field, while the smoothed $\lambda_1'$ and $\lambda_2'$ maps are given in the centre and bottom panel, respectively.  The grayscale map of $\lambda_1'$ gives a strong impression of filamentary structure.  Note, however, that the $\lambda$ parameters are essentially edge-finders and both overdensities (whose edges are apparent in the $\lambda_1'$ map) and underdensities ($\lambda_2'$ map) will have boundaries.  In a Gaussian random field, the overdensities and underdensities are statistically identical, so the maps are morphologically similar.

On the righthand side of Fig.~\ref{fig:GRFCompare2d_6}, we show the
same grayscale maps for the $z=0$ dark matter distribution in a cosmological
simulation (see \S~\ref{sec:Sims}), but with the $\lambda$
parameters (centre and bottom panels) normalized to {\it local}
density\footnote[2]{Note that the gaussian random fields used in the
left columns of Figs.~\ref{fig:GRFCompare2d_6} and
\ref{fig:CombinedContours} are {\it not} the seed fields for the
simulations use in the right columns.}.  Non-linear growth of
structure amplifies overdense edges and the low-contrast structures
tend to become washed out in the raw $\lambda$-maps.  After density
normalization, however, the edges show a similar morphology to those
in the $\lambda_1'$ and $\lambda_2'$ maps of the gaussian random
field, suggesting that the outline of the network of filaments and
walls is already in place in the initial conditions (as was
first suggested by, Bond et~al. 1996).  Also included in the bottom two
rows of Fig.~\ref{fig:GRFCompare2d_6} are bars indicating the
direction of the local axis of structure.  In the simulations, the
axis of structure is nearly perfectly aligned with the $\lambda_1'$
edges in even the weakest strands, while in the gaussian random
fields, there is no obvious correlation between the axis of structure
and the orientation of the edges.  The direction of the axis of
structure does not appear to be correlated on much more than a
smoothing length in the gaussian random field, while the simulations
have structures that are aligned across the entire $200$\hmpc~box!

\section{Geometry of dark matter and mock galaxy distributions}
\label{sec:DSpace}

\begin{figure}[t]
\epsscale{0.8}
\plotone{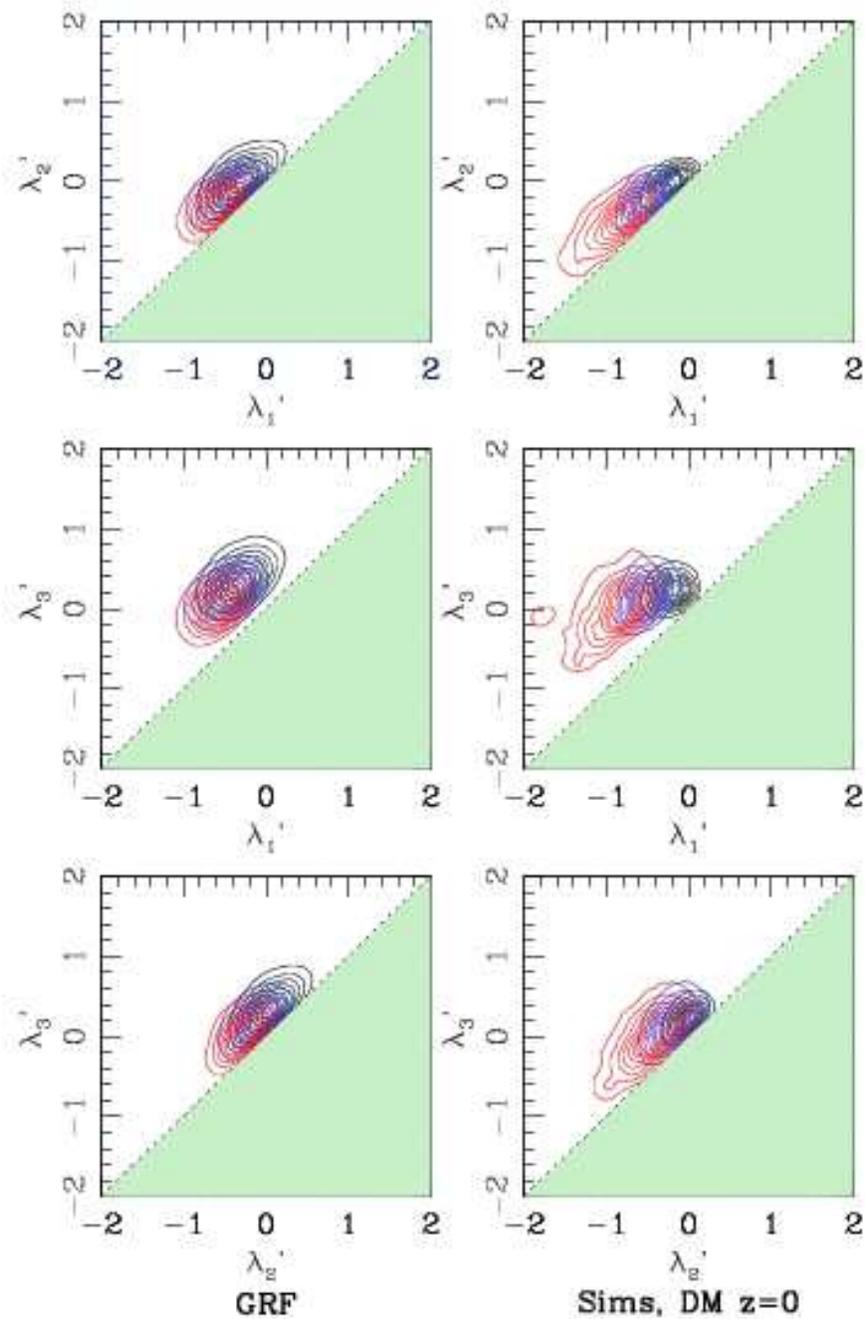}
\caption{The $\lambda$-space projections for a three-dimensional
gaussian random field (left) and the simulated $z=0$ dark matter
distribution (centre).  Contours represent the distribution of
$\lambda$ at the positions of dark matter particles (not grid cells)
from a sampling of the fields.  In the left two columns, contours are
for different density subsamples, where black contours indicate the
distribution of points in underdense regions (first or second
quartile), blue contours in regions of intermediate density (third
quartile), and red contours in the most dense regions (upper
quartile).  \label{fig:CombinedContours}}
\end{figure}

The two-point correlation function and power spectrum already provide a direct measure of the two-point properties of the large-scale distribution of matter, so we wish to extract from $\lambda$-space only that information which cannot be obtained from two-point statistics.  In the left column of Fig.~\ref{fig:CombinedContours}, we show the $\lambda$-space distribution (smoothed on a $10$\hmpc~scale) of a three-dimensional gaussian random field.  The contours illustrate $\lambda$-space distributions for points in the bottom $50$ per cent of the real-space density distribution (black), between $50$ and $75$ per cent (blue), and the upper $25$ per cent (red).  The gaussian random field was sampled in a ($200$~Mpc)$^3$ three-dimensional grid in proportion to the local overdensity, which was normalized to match the amplitude of the $\Lambda CDM$ non-linear power spectrum at each smoothing scale.  No points were placed in grid cells having $\delta\le -1$ ($<0.1$\% of the volume in all cases presented here).  Within a given panel of Fig.~\ref{fig:CombinedContours}, there is little variation in the size or shape of the contours as a function of overdensity, indicating no density dependence of structural morphology.  The plot is generated from only one realization of a gaussian random field, but finite volume effects are small at this smoothing scale.

In the large-scale distribution of matter, structure grows due to the
gravitational effects of the total matter distribution.  As such, we
compute $\lambda$-space distributions at the positions of {\it dark
matter particles} (not grid cells) in the $z=0$, dark-matter-only
simulation described in \S~\ref{sec:Sims}; these are shown in the
right column of Fig.~\ref{fig:CombinedContours}.  At high densities
(red), the dark matter contours are extended because of the presence
of overdense clumps.  The low-density contours, however, are
contracted relative to those of the gaussian random field.  The
largest underdensities in the gaussian random field have a positive
curvature (and, therefore, positive $\lambda$-values) comparable to
the negative curvature of the overdensities, while dark matter
underdensities are voids with a very small spread in density and
curvature.  If non-linear growth of structure leads to the formation
of filaments and/or walls in the dark matter distribution, there
should be an excess in the regions of $\lambda$-space occupied by
these fingerprints.

\begin{figure}[t]
\epsscale{1.0}
\plotone{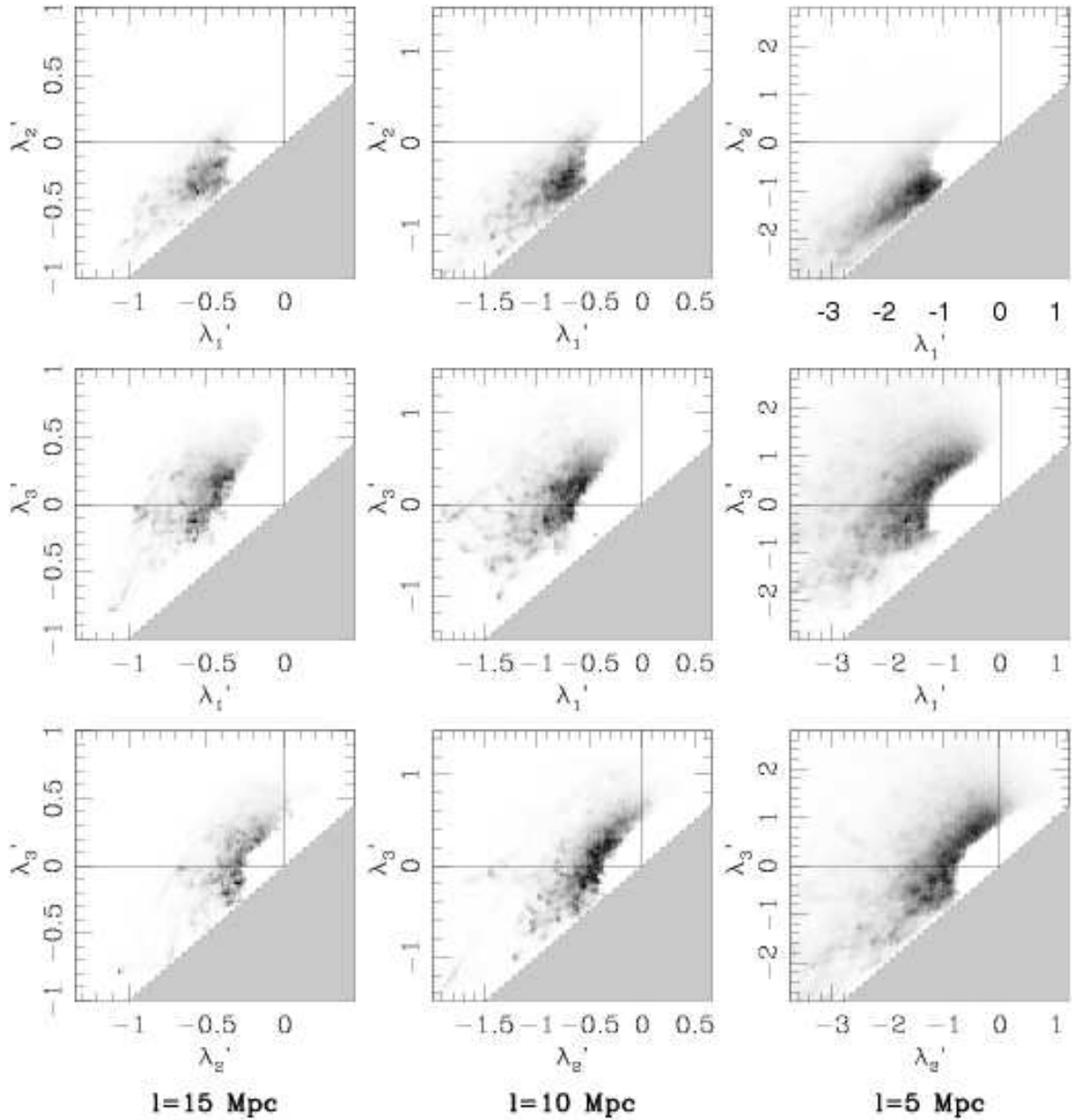}
\caption{The $\lambda$-space projections for the $z=0$ dark matter
distribution, after subtraction by the $\lambda$-space distribution of a
gaussian random field with the same power spectrum.  Grayscale cells
indicate those regions in which the dark matter distribution shows an
excess over the gaussian random field and for which the local density
is greater than the mean, with darker cells indicating more of an
excess.  Regions showing no excess or regions showing a deficit are
left blank.  The axes of each plot have been scaled in proportion to
the amplitude of the dimensionless power spectrum at the appropriate
smoothing scale.\label{fig:EvolveContours_z0}}
\end{figure}

In order to bring out these regions,
Fig.~\ref{fig:EvolveContours_z0} shows the $\lambda$-space
distributions of dark matter after subtracting a gaussian random field
with the same power spectrum (i.e. the difference between the columns
in Fig.~\ref{fig:CombinedContours}).  Here, grayscale density maps
replace contours and there is no colour-coding for real-space
density, but only points with density greater than the mean are
included.  A comparison of these plots with the fingerprints in
Fig.~\ref{fig:OneObjBox3} will help us determine which structures
dominate the dark matter distribution at each smoothing scale.  Note
that the axes of each plot have been scaled in proportion to the
amplitude of the dimensionless power spectrum at the appropriate
smoothing scale, allowing for a direct comparison of structural
morphology between scales.

In order to make this comparison more quantitative, we will define a set of structural parameters,
\begin{equation}
R^{\lambda}_{{\rm ij}}=\overline{1-\frac{\lambda_{\rm j}}{\lambda_{\rm i}}},
\end{equation}
where the mean is computed over the region of $\lambda$-space having $\lambda_{\rm i}<0$ and $\lambda_{\rm j}<0$.  For example, $R^{\lambda}_{12}$ quantifies the mean axis ratio for structures that are negatively curved along at least two principal axes of the Hessian ellipse, allowing one to distinguish walls from filaments.  A small value of $R^{\lambda}_{12}$ indicates that the axes have similar curvature and are more like filaments, while $R^{\lambda}_{12}\sim 1$ indicates the dominance of wall-like structures.  Similarly, $R^{\lambda}_{23}$ can be used to distinguish filaments from clumps in regions with negative curvature along all three principal axes.

As shown in the righthand column of Fig.~\ref{fig:OneObjBox3}, the densest parts of clumps lie on the lower boundary of $\lambda$-space (dashed line) in all three projections.  Filaments concentrate on the lower boundary of the $\lambda_2'$--$\lambda_1'$ projection, as do clumps, but filaments also concentrate along the x axis of the other two projections.  In Fig.~\ref{fig:EvolveContours_z0}, we see that at both $15$ and $10$\hmpc~scales, the points in Fig.~\ref{fig:EvolveContours_z0} appear to be shifted up toward the x axis relative to the clump fingerprint.  The distribution of points in the $\lambda_3'$ vs.~$\lambda_1'$ projection is broad and extends to very negative values of $\lambda_1'$, consistent with the presence of filaments in addition to clumps.  At $z=0$, it appears that filaments are the dominant over clumps at $15$\hmpc~scales ($R^{\lambda}_{23}=0.605$) and $10$\hmpc~scales ($R^{\lambda}_{23}=0.602$), with clumps becoming more prominent at $5$\hmpc~smoothing scales ($R^{\lambda}_{23}=0.554$).  This evolution is most apparent in the highest contrast structures (small $\lambda_2$).

There is little evidence for walls at $z=0$ at the smoothing scales shown.  Walls manifest themselves as an excess of points near both $\lambda_2'=0$ and $\lambda_3'=0$.  Although they are difficult to distinguish from filaments in the $\lambda_3'$ vs.~$\lambda_1'$ projection, there would be a clear overdensity of points along the x axis of the top plots of Fig.~\ref{fig:CombinedContours} and near the origin of the bottom plots if walls were dominant features.  The mean eigenvalue ratios suggest that structures become less wall-like at $5$\hmpc~smoothing scales, with $R^{\lambda}_{12}=0.370$, as compared to $R^{\lambda}_{12}=0.485$ at $15$\hmpc~smoothing scales.

\begin{figure}[t]
\plotone{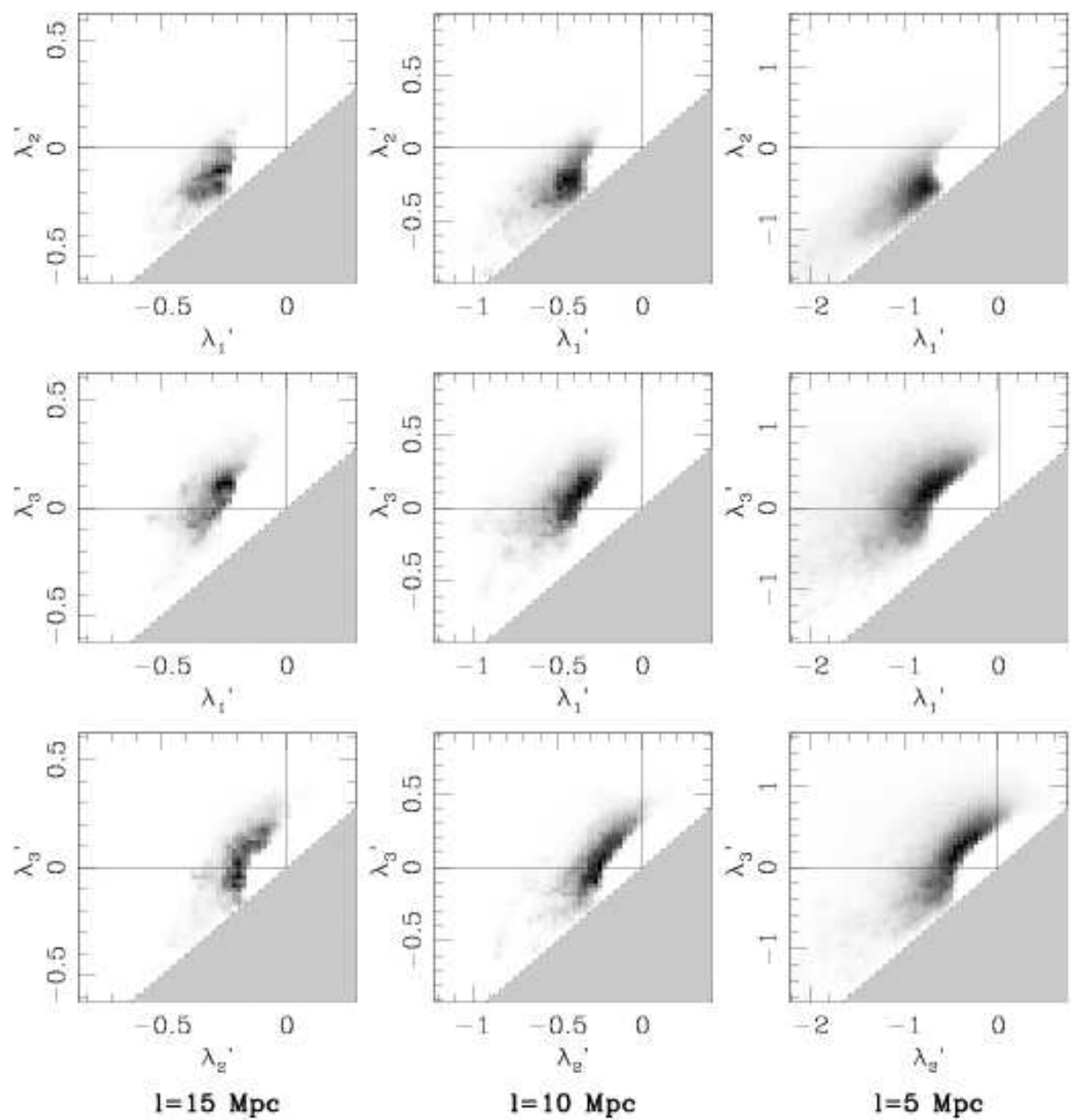}
\caption{The same as Fig.~\ref{fig:EvolveContours_z0}, but for the
$z=1$ dark matter distribution.\label{fig:EvolveContours_z1}}
\end{figure}

\begin{figure}[t]
\plotone{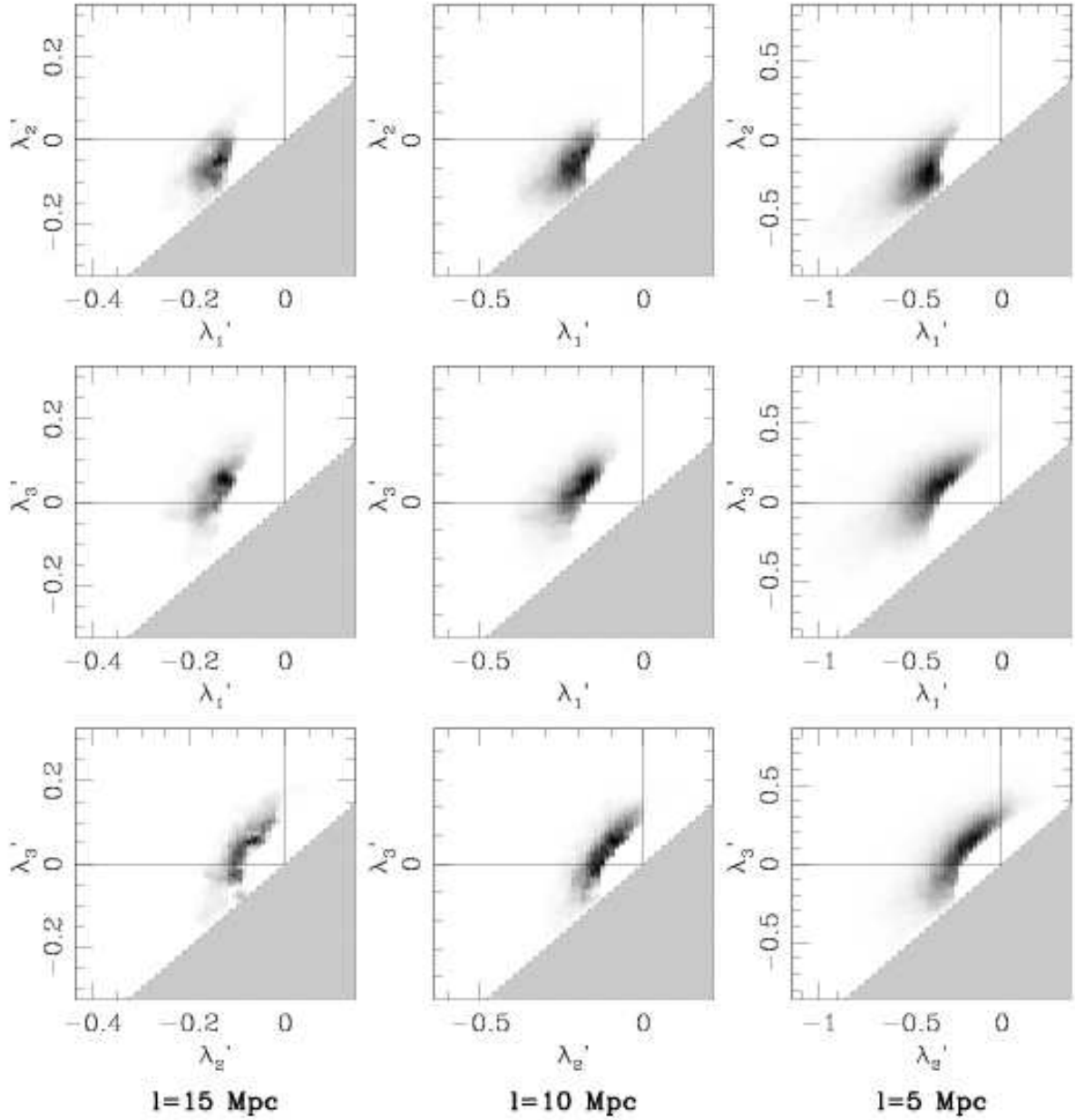}
\caption{The same as Fig.~\ref{fig:EvolveContours_z0}, but for the
$z=3$ dark matter distribution.  At $15$\hmpc~and $10$\hmpc~scales,
distributions appear more wall-like than those at $z=0$, suggested by
a shift towards the $\lambda_2=0$ axis in the $\lambda_2$--$\lambda_1$
projection.  Meanwhile, the $\lambda$-space distribution at
$5$\hmpc~scales is more filamentary, as suggested by a similar shift
toward the $\lambda_3=0$ axis in the bottom two projections.  These
latter projections are very similar to those for $15$\hmpc~scales at
$z=0$.\label{fig:EvolveContours_z3}}
\end{figure}

In Figs.~\ref{fig:EvolveContours_z1} and \ref{fig:EvolveContours_z3}, we present the $\lambda$-space distribution of structures at $z=1$ and $z=3$, respectively, smoothing on the same three comoving length scales as in Fig.~\ref{fig:EvolveContours_z0} ($z=0$).  The morphology of overdense structures in the dark matter distribution changes significantly at these redshifts.  In all three $\lambda$-space projections, the distribution is concentrated closer to the x axis at higher redshift, meaning that the larger eigenvalues decrease more rapidly than the smaller ones and overdensities are approaching spherical symmetry.

At $l=15$\hmpc, all of the plots are more sparsely populated at high redshift, indicating that the $\lambda$-space distribution is closer to that of a gaussian random field.  This is especially true of $\lambda_3'$ vs.~$\lambda_2'$, which also shows no dramatic change in shape out to $z=1$.  In the $\lambda_3'$ vs.~$\lambda_1'$, $l=15$\hmpc~projection, the most densely populated regions retain their shape and relative contrast out to the highest redshift ($R^{\lambda}_{13}=0.794$ at $z=3$ as compared to $R^{\lambda}_{13}=0.764$ at $z=0$), but some of the sparse features in the bottom left corner fade by $z\sim 1$.  Finally, the upper left panel, $\lambda_2'$ vs.~$\lambda_1'$, undergoes strong evolution, with the tails at very negative eigenvalues disappearing and the bulk of the distribution moving to the $\lambda_2'=0$ axis.  This evolution is also demonstrated in $R^{\lambda}_{12}$, which drops from $0.599$ at $z=3$ to $0.485$ at $z=0$.

In Fig.~\ref{fig:OneObjBox3}, we see that walls are most easily distinguishable from other structures in $\lambda_2'$ vs.~$\lambda_1'$, with points concentrated along $\lambda_2'=0$.  Taken together, these facts are suggestive of wall-to-filament evolution at a $l=15$\hmpc~comoving smoothing scale.  The rapid fading of the $\lambda_3'$ vs.~$\lambda_2'$ map is also explained, as walls are concentrated toward the origin in these two eigenvalues and would be indistinguishable from a gaussian random field.  The signature of walls at high redshift are apparent in these plots, but they are of low contrast compared to the structures seen at smaller scales and lower redshift.

The two smaller smoothing lengths, $l=10$\hmpc~and $l=5$\hmpc, exhibit a filament-to-clump evolution.  In $\lambda_3'$
vs.~$\lambda_2'$, the distributions at both length scales exhibit a
tail along the dashed boundary that grows toward $z=0$.  This tail is
populated by the points near the dense centres of clumps.  A similar
tail exists in $\lambda_3'$ vs.~$\lambda_1'$, but here the growth of
individual filamentary features is also evident, with a concentration
near the $\lambda_3'=0$ axis.  The eigenvalue ratios, $R^{\lambda}_{23}=0.647$ at $z=3$ and $R^{\lambda}_{23}=0.554$ at $z=0$, suggest filament dominance at $l=5$\hmpc~across the entire redshift range, but with an evolution towards more ``clump-like'' filaments.

It is interesting to compare the $z=1$ projections for $l=10$\hmpc~to the $z=0$, $l=15$\hmpc~projections.  Their similarity suggests that the two length scales are in similar stages of non-linear evolution, but at different times.  This conclusion is supported by the eigenvalue ratios, with $l=15$\hmpc~having $R^{\lambda}_{12}=0.485$ at $z=0$ and $l=10$\hmpc~having $R^{\lambda}_{12}=0.497$ at $z=1$. That the smaller length scales develop filamentary structure earlier is an indication of bottom-up structure formation; however, at a given comoving scale, structure evolves from walls to filaments to clumps, consistent with the ellipsoidal collapse picture of \citet{Pancake}.

\subsection{ Mock catalogues in $\lambda$-space }
\label{subsubsec:GalSimDspace}

In the SDSS data, we will be working with the observed galaxy
distribution, so it is useful to generate $\lambda$-space projections
for mock galaxies, generated by the methods described in
\S~\ref{sec:Sims}.  There are many fewer galaxies than dark matter
particles in the box volume, so the signal will be much weaker and
shot noise may be non-negligible at $l=5$--$10$\hmpc~scales.  More
importantly, there may be interesting differences between the
large-scale structures appearing in the dark matter and galaxy
distributions.  In order to populate the dark matter haloes with
galaxies, we must know the relevant halo occupation distribution
\citep[HOD,][]{HOD}; the probability, $P(N|M)$, that a halo of mass
$M$ will contain $N$ galaxies.

Here we use the parametrization of \citet{HODPars2},
in which the probability of a halo of mass $M$ containing a central
galaxy is given by:
\begin{equation}
\langle N_{\rm cen}\rangle=\frac{1}{2}\left[1+{\rm erf}\left(\frac{{\rm log} M - {\rm log} M_{\rm min}}{\sigma_{{\rm log}~M}}\right)\right].
\label{eq:ncen}
\end{equation}
Each central galaxy may be surrounded by one or more satellite
galaxies, the mean number of which is given by,
\begin{equation}
\langle N_s\rangle =\frac{1}{2}\left[1+{\rm erf}\left(\frac{{\rm log} M - {\rm log} M_{\rm min}}{\sigma_{{\rm log}~M}}\right)\right]\left(\frac{M-M_0}{M_1}\right)^{\alpha}.
\label{eq:ns}
\end{equation}
The number of satellite galaxies in a given halo is drawn from a
Poisson distribution and has no upper limit.  The free parameters,
$M_{\rm min}$, $\sigma_{{\rm log}~M}$, $M_0$, $M_1$, and $\alpha$ have been
derived from fits to SDSS galaxy catalogues.  Table $1$ of
Zheng et~al. 2007 provides the best-fitting HOD parameters for a series of
absolute $r$-band magnitude cutoffs.

Once drawn from the relevant distributions, central galaxies are given
the mean position and peculiar velocity of their associated dark matter halo,
while satellite galaxies are each given the position and velocity of a
randomly selected dark matter particle within the halo.  When working
with the SDSS data, we will construct a series of volume-limited
samples, so the mock catalogues must be prepared to mimic the associated
selection function. 

One important difference between the redshift-space galaxy
distribution and the dark matter distribution is the presence of
fingers-of-god.  These features, created by the peculiar velocities
of galaxies in clusters, appear as very narrow filaments along the
line of sight with widths $\sim 1$\hmpc~and lengths up to $\sim
10$\hmpc.  The mock catalogues must contain a model of these redshift
distortions if they are to properly mimic the SDSS data.  The
`distorted' positions of galaxies are computed,
\begin{equation}
\boldsymbol{r}_z=\boldsymbol{r}+\frac{\boldsymbol{r}\mathbf{\cdot}\boldsymbol{v}}{H_0}\boldsymbol{\hat{r}},
\label{eq:zdistort2} 
\end{equation} 
where $\boldsymbol{r}$ is the position in real space, $\boldsymbol{v}$ is the
galaxy's peculiar velocity, and $\boldsymbol{\hat{r}}$ is the unit radial
vector between the observer and the galaxy.  

\begin{figure}[t]
\plotone{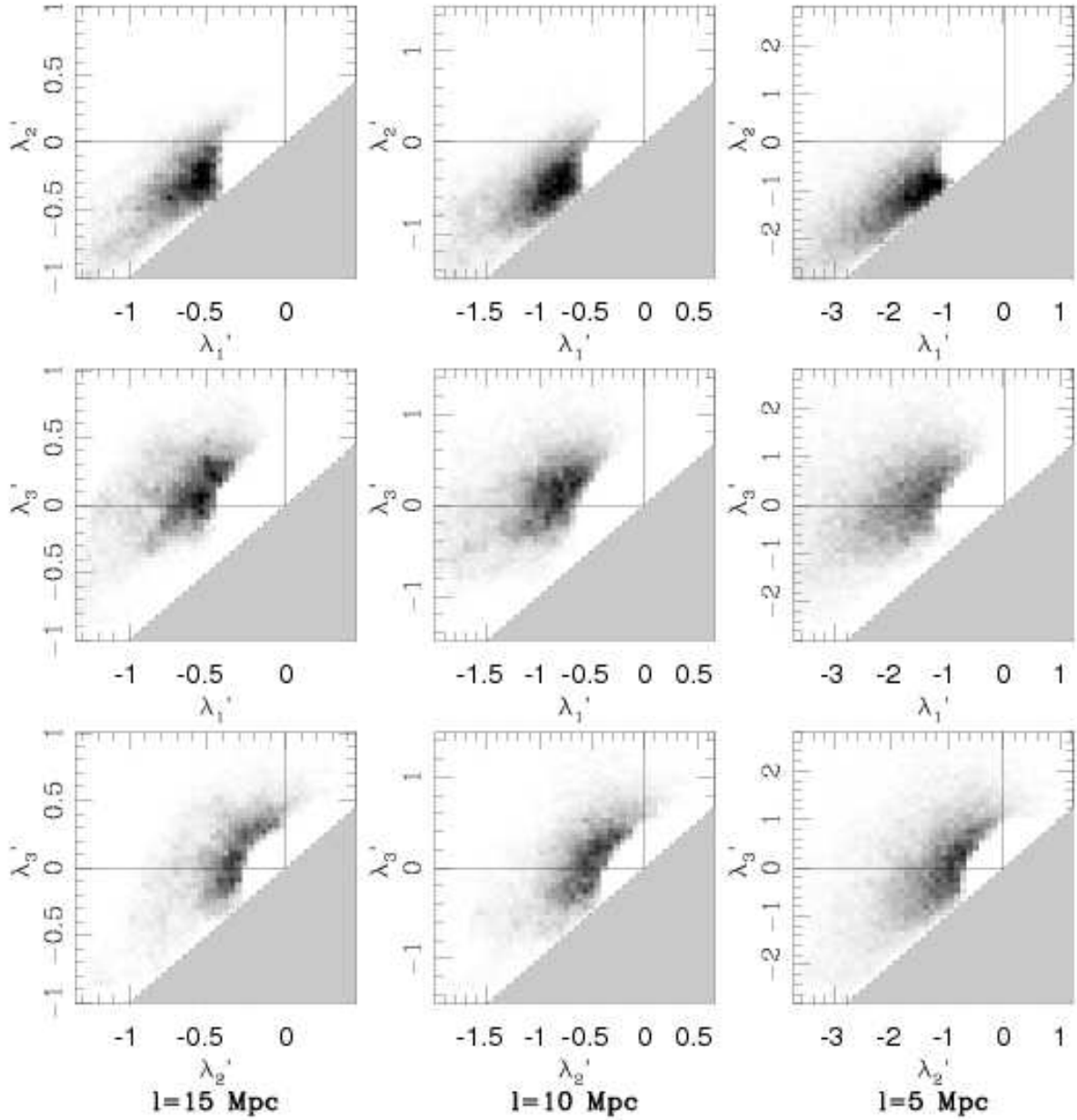}
\caption{The $\lambda$-space projections for a $z=0$ mock galaxy catalogue
with $M_r<-20.5$, including redshift
distortions. \label{fig:ZGalContours}}
\end{figure}

In Fig.~\ref{fig:ZGalContours}, we plot the $\lambda$-space projections of an $M_r<-20.5$ mock galaxy sample projected into redshift space.  The redshift distortions are applied from a vantage point with coordinates \{-300,0,100\}\hmpc~relative to the origin of the box, selected to closely resemble the data samples constructed in the next section.  The morphological characteristics of the $\lambda$-space distributions for $l=15$ and $10$\hmpc~are very similar to those of Fig.~\ref{fig:EvolveContours_z0}, though several features are different (for example, the smearing of the previously noted clump signature in the bottom left corner of $\lambda_3'$ vs.~$\lambda_1'$).  The changes are most noticeable at $l=5$\hmpc, where all of the $R^{\lambda}$ increase by $\sim 0.05$ in the mock catalogs and the high-contrast features all but disappear.  These high-contrast features are primarily due to clusters, which become lower-contrast fingers-of-god when projected into redshift space. 

With the construction of the mock catalogues and the development of the
theoretical framework presented in \S~\ref{sec:Method}, we now have
all of the tools necessary to compare the real galaxy distribution to
that predicted by the standard cosmological model.  In the next
section, we will present our galaxy sample, drawn from the SDSS, and use
the $\lambda$-space distribution to make inferences about real
large-scale structure.

\section{ Filaments and walls in the SDSS galaxy distribution }
\label{sec:Data}

\begin{figure}[t]
\plotone{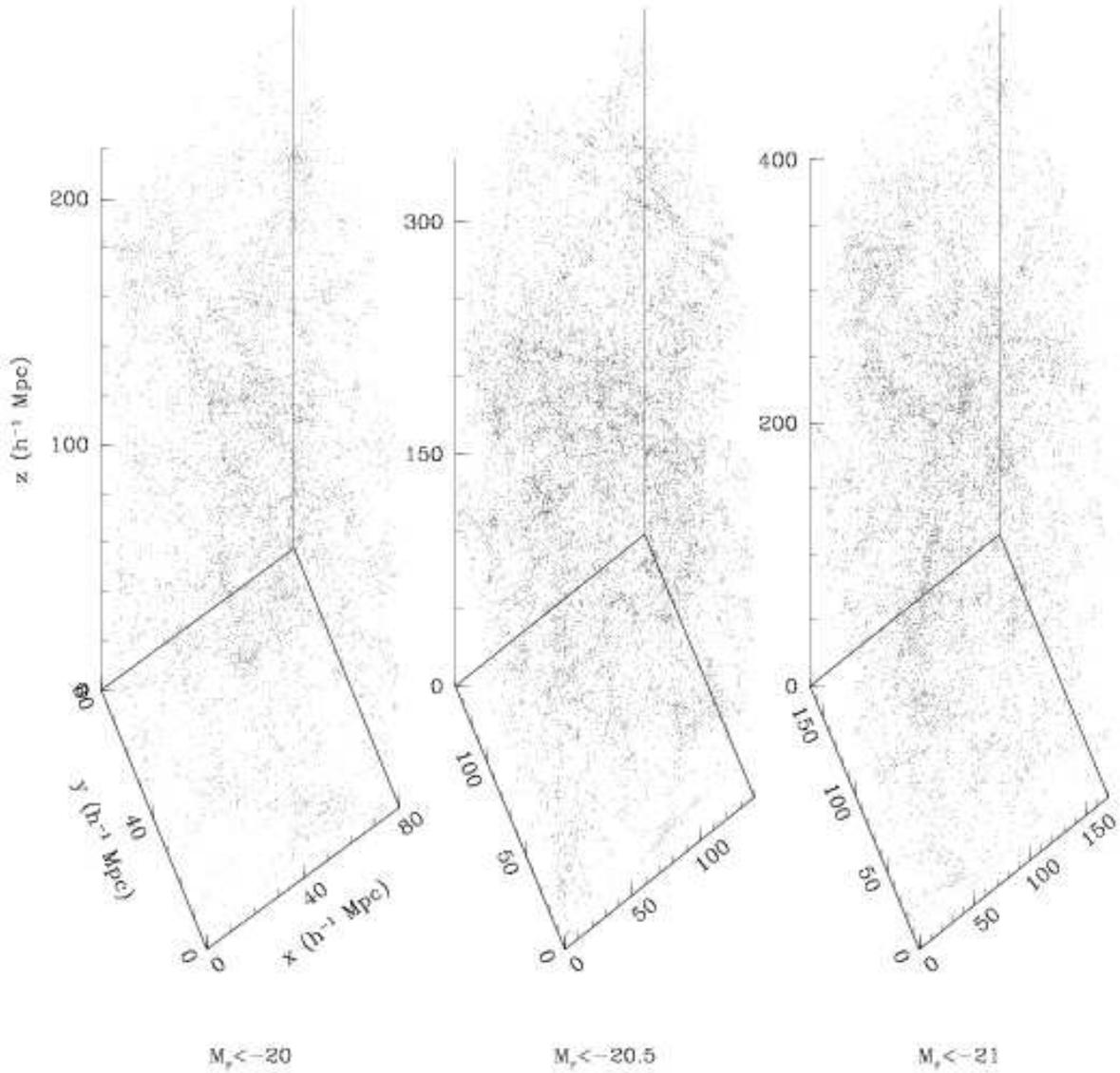}
\caption{The three volume-limited samples taken from the SDSS VAGC
large-scale structure sample, with galaxies placed at their comoving
positions based on the concordance cosmology.  The location of the
Milky Way is ${\boldsymbol r}=(250,-20,110)$\hmpc,
${\boldsymbol r}=(310,-20,170)$\hmpc, and ${\boldsymbol r}=(400,-25,200)$\hmpc~in the
Mr$20$, Mr$205$, and Mr$21$ samples, respectively, and the z axes
are parallel to the Galactic North pole.\label{fig:DataBoxes}}
\end{figure}

The Sloan Digital Sky Survey has imaged a quarter of the sky in five wavebands, ranging from $3000$ to $10000$~\AA, to a depth of $r \sim 22.5$ \citep{DR7}.  As of Data Release $7$, spectra have been taken of $\sim 930,000$ galaxies, covering $8032$ deg$^2$ and extending to Petrosian $r \sim 17.7$ \citep{Spectro}.  Galaxy redshifts are typically accurate to $\sim 30$~km~s$^{-1}$, making the survey ideal for studies of large-scale structure.  For this study, we need a portion of sky with relatively few coverage gaps to minimize the effect of the window function on the $\lambda$-space distributions.  With this in mind, we construct three volume-limited subsamples from the northern portion ($8<\alpha<16$~h and $25<\delta<60$) of the NYU Value-Added Galaxy Catalogue \citep[NYU-VAGC, ][, through DR$6$]{VAGC}, the first $80 \times 80 \times 220$~($h^{-1}$~Mpc)$^3$~in size with $M_r<-20$ (Mr$20$), another $140 \times 140 \times 340$~($h^{-1}$~Mpc)$^3$~in size with $M_r<-20.5$ (Mr$205$), and finally $170 \times 170 \times 400$~($h^{-1}$~Mpc)$^3$~in size with $M_r<-21$ (Mr$21$).  Absolute magnitudes were computed with {\it kcorrect} using SDSS Petrosian magnitudes shifted to $z=0.1$ (and using $h=1$).  The sample properties of the galaxy distribution are given in Table~\ref{tab:DataSamples} and the galaxy distributions are plotted in Fig.~\ref{fig:DataBoxes}.

All boxes are padded with $30$\hmpc~of empty space and binned on a
$128 \times 128 \times 256$ grid, resulting in grid spacings of $1.09$,
$1.56$, and $1.80$\hmpc~in the Mr$20$, Mr$205$, and Mr$21$ subsamples,
respectively.  To correct for edge effects, we define a window
function on the same grid such that,
\begin{equation}
\label{eq:Window}
W_{\rm i}=\left\{
  \begin{array}{lll}
     1 & \textrm {{\it if}} & {\rm i}\in S \\
     0 & \textrm {{\it if}} & {\rm i} \notin S
  \end{array}\right.
\end{equation}
where $S$ is the set of grid points within the SDSS survey volume and
$i$ is a grid cell within the subsample volume.  The above window
function was computed by smoothing the sample boxes on a
$20$\hmpc~scale and including in the window only those regions with
densities more than $4$ standard deviations below the mean density in
the box.  This is a crude approximation, but sample boxes were
selected to lie in regions with few SDSS coverage gaps, so only a
small fraction of grid cells are excluded (primarily near the box
boundaries).

Once the window function and density fields have been computed on the
grid, we smooth both with gaussian kernels on the scales given in
Table~\ref{tab:DataSamples}, giving $\tilde{W}({\boldsymbol x})$ and
$\tilde{\rho}({\boldsymbol x})$, respectively.  The corrected density field is then
\begin{equation}
\label{eq:DensityCorrect}
{\rho'}({\boldsymbol x})=\frac{\tilde{\rho}({\boldsymbol x})}{\tilde{W}({\boldsymbol x})}.
\end{equation}

The first and second derivatives of the density field have zero mean
on large scales (if the cosmological principle holds) and the kernels
used to compute them will integrate to zero.  As such, there is no
need to {\it renormalize} the smoothed derivative fields near the
padded regions.  Rather, we need only subtract the signal from the
edge artificially introduced by the padding.  That is,
\begin{equation}
\label{HessCorrect}
{H'}_{\rm ij}=\tilde{H}_{\rm ij}-\bar{\rho}\tilde{B}_{\rm ij},
\end{equation}
where $\tilde{B}_{\rm ij}$ is the matrix of smoothed second partial
derivatives of the window function.

For each of the data samples listed in Table~\ref{tab:DataSamples}, we
constructed a mock galaxy catalogue that matched the box dimensions,
number density, and padding of
the SDSS data, incorporating redshift distortions accordings to the prescription described in \S~\ref{subsubsec:GalSimDspace}.  The processing on the mock galaxy catalogues and the real data
were identical and they should be directly comparable, with the caveat
that the mock catalogues were taken from a $200 \times 200 \times
200$(\hmpc)$^3$~box periodic box, and therefore had to be repeated along the
z axis in order to match the dimensions of the data subsamples.  As
such, finite-volume uncertainties will be larger in the simulations.

\begin{figure}[t]
\plotone{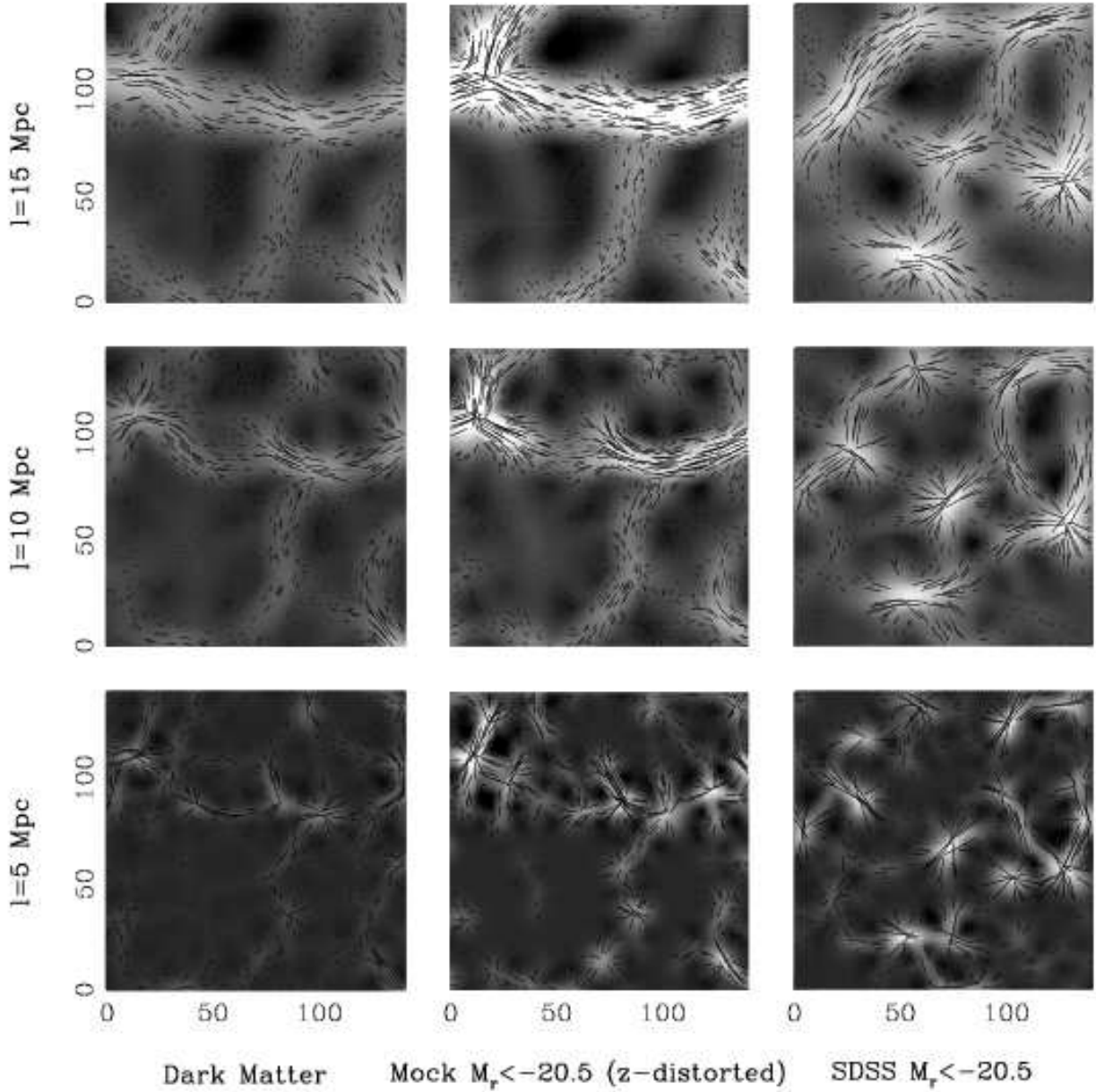}
\caption{
The two-dimensional $\lambda_1'$ distributions in slices from the
$z=0$ output of a {\lcdm} cosmological simulation (dark matter and
mock galaxy distributions, left and centre columns), and the SDSS
galaxy distribution (right column), smoothed on a $15$\hmpc~scale (top
row), a $10$\hmpc~scale (middle row) and a $5$\hmpc~scale (bottom
row). Bars have been plotted over the maps to indicate the direction
of the axis of structure at each point in space.  In all cases, the
bar's length is proportional to the magnitude of $\lambda_1'$ at that
point.  The grayscales are identical between the three columns.  In
all three samples and on all three smoothing scales, the axis of
structure follows the minima in $\lambda_1'$.
\label{fig:DataCompare3}}
\end{figure}

In Fig.~\ref{fig:DataCompare3}, we compare the two-dimensional
$\lambda_1'$ maps between $10$\hmpc~thick slices from the $M_r<-20.5$
subsample, its corresponding mock catalogue, and the dark matter
distribution.  These data represent $\sim 3$ per cent of the $M_r<-20.5$
subsample and the normalization and range of the grayscales are
identical between the two columns of each figure.  The mean properties
of each slice are superficially similar, although the slice from the
mock catalogue has a large filament running through its centre,
dominating the visual impression.  In all maps, the axis of structure
is oriented along edges in $\lambda_1'$, unlike the gaussian random
field shown in Fig.~\ref{fig:GRFCompare2d_6}.

\begin{figure}[t]
\plotone{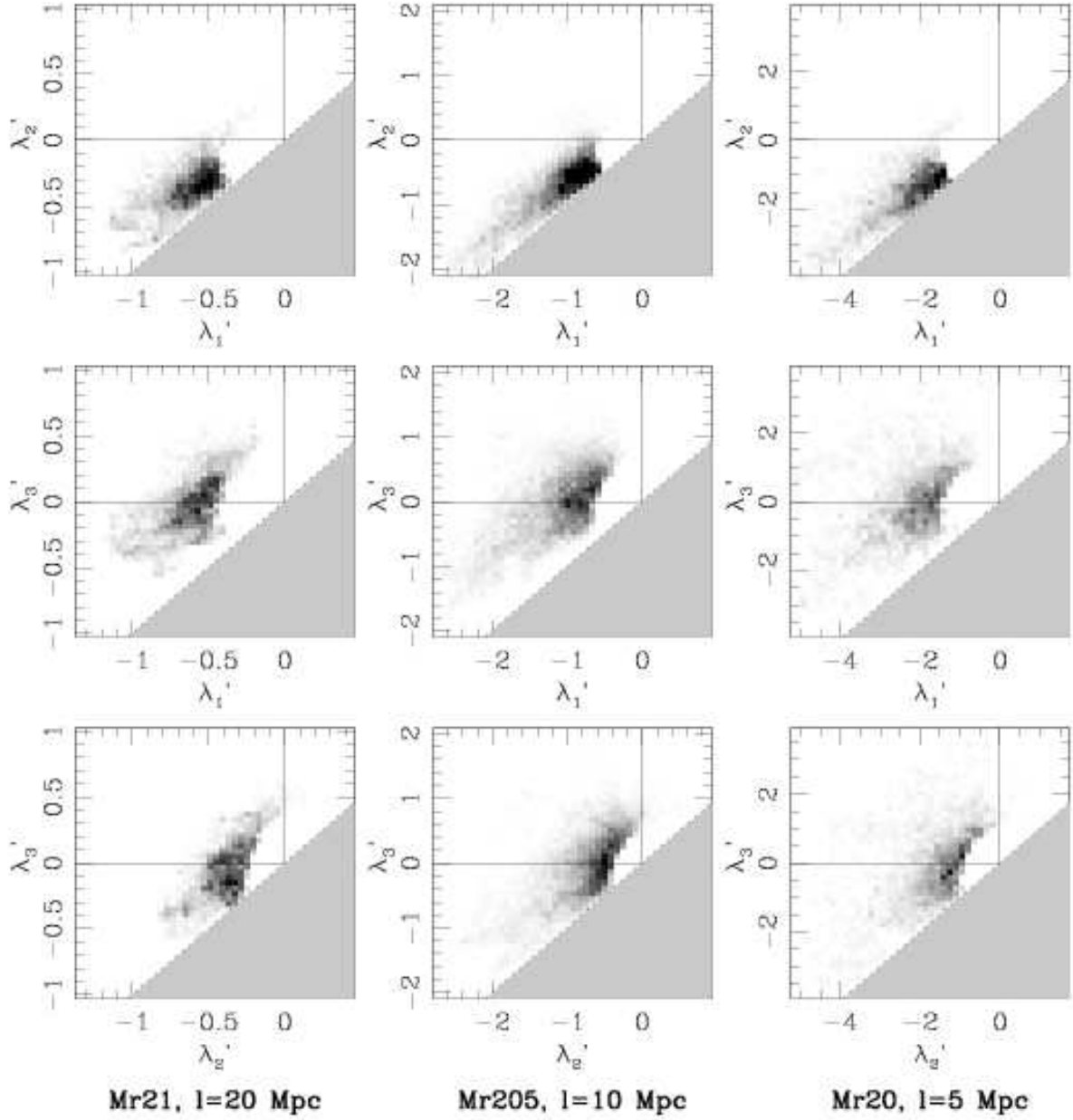}
\caption{ The $\lambda$-space projections for the three
volume-limited subsamples of galaxies in the VAGC galaxy catalogue,
smoothed on $20$, $10$, and $5$\hmpc~scales.  Grayscale maps are the
same as in Fig.~\ref{fig:EvolveContours_z0}.\label{fig:DataContours4}}
\end{figure}

\begin{figure}[t]
\plotone{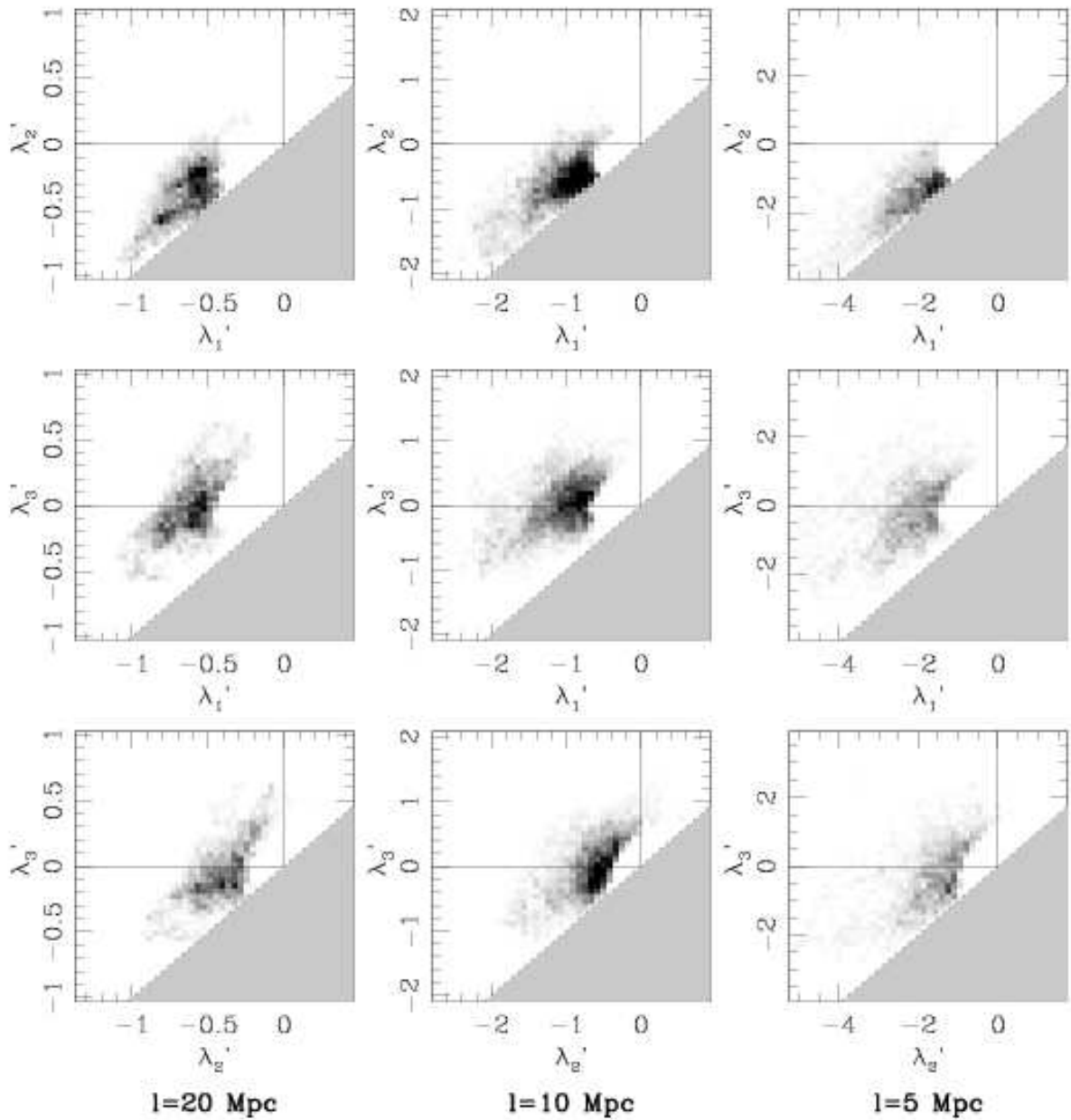}
\caption{The same as Fig.~\ref{fig:DataContours4}, but for the mock galaxy samples.  Grayscales and axis limits are the same between the two figures.\label{fig:MockContours4}}
\end{figure}

The $\lambda$-space projections of the SDSS data are displayed in Fig.~\ref{fig:DataContours4} for $l=20$\hmpc~(Mr$21$), $l=10$\hmpc~ (Mr$205$), and $l=5$\hmpc~(Mr$21$).  For comparison, we present the corresponding $\lambda$-space distributions for the redshift space mock catalogues in Figs.~\ref{fig:MockContours4}.  At small scales, the value of $\lambda_1'/\lambda_2'$ at the positions of galaxies is generally near unity (diagonal lines) and the distribution morphologies are similar to those in the mock catalogues.  Shot noise is more severe at these scales than at larger scales, so the similarity could be in part due to the extra power added by Poisson fluctuations.  On $10$\hmpc~scales, there are some indications that the real galaxy distribution may be more evolved than the simulated one, with all $R^{\lambda}$ being smaller in the data by $\sim 0.05$.  The $\lambda$-space distributions for the data appear to have more high-contrast structures, including a tail of clump-like objects with near-unity values of $\lambda_2'/\lambda_3'$.  If real, the discrepancy could suggest that the simulation is underestimating $\sigma_8$.

Finite-volume effects are apparent at the larger smoothing scales for all of the samples, where the $R^{\lambda}$-values are sometimes larger and sometimes smaller than their counterparts in the mock catalogues.  At the largest smoothing scales, there is clear evidence for non-linearity and the $\lambda_2'$--$\lambda_1'$ projection shows evidence for both filaments and walls.  Unlike the distributions at $l=5$\hmpc~and $l=10$\hmpc, the data do not appear to be much more clump-like than the mock catalogues ($R^{\lambda}_{23}=0.61$ in the data and $R^{\lambda}_{23}=0.54$ in the mock catalogues).  However, there is some suggestion that the data are {\it less} wall-like on the largest scales ($R^{\lambda}_{12}$ is smaller in the data by $\sim 0.08$).  This would also be consistent with a more evolved population, though finite-volume effects and uncertainties in the halo occupation distribution make it difficult to say for sure.

\section{Discussion}
\label{sec:Discussion}
\subsection{A matter of scale}
\label{subsubsec:Scale}

Mention of large-scale structure often brings to mind images of
narrow, interconnected filaments with clusters at the interstices
weaving their way between vast cosmic voids.  Bond et~al. 1996 dubbed
this mental image the `cosmic web', but we have avoided using that
phrase in this paper.  The components of a spider web are strands,
which come together at nodes to make a variety of complex structures
on larger scales.  The same could be said for filaments in the dark
matter distribution, but that is not a complete picture.  We also know
that these filaments are made up of bound dark matter haloes.  These
small-scale filaments can themselves be the building blocks for
large-scale ones.

Some recent attempts to quantify and identify individual filamentary
structures have not taken the multiscale nature of structure into
account.  For example, minimal spanning tree algorithms
\citep[e.g.,][]{MST85,MST07} attempt to select filaments in the
same manner as clusters, assuming that a step down up in linking
length (that is, down in density contrast) will allow identification
of the strands and full reconstruction of the web.  Adaptive smoothing
kernels and tesselation methods \citep[e.g.,][]{Delaunay} enhance
high-density regions in much the same way that simulations of galaxy
formation adapt the mesh to treat star-forming regions, resulting in
a nebulous combination of structure on small scales and large scales.
There have also been attempts to use image segmentation \citep{MMF} to
select the most `prominent' large-scale structures across a
multitude of scales, and then remove them before looking for the next
ones.  Although this procedure makes sense {\it on a given scale},
removal of structures on one scale can cause one to miss structures on
another.

\begin{figure}[t]
\epsscale{0.4}
\plotone{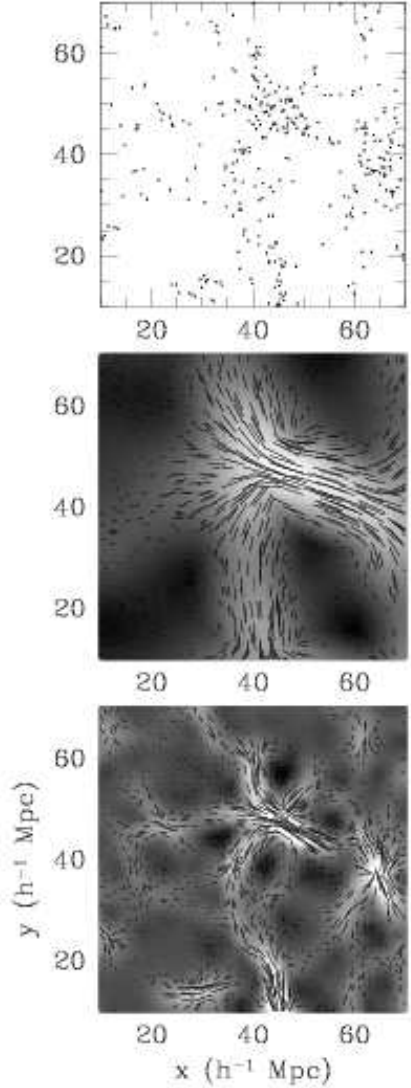}
\caption{
A two-dimensional demonstration of the structural hierarchy in a
$15$\hmpc~deep slice from the SDSS Mr$20$ subsample.  Galaxies are
plotted in the upper left panel and $\lambda_1'$ grayscale maps are
shown in the other two panels, smoothed on a $10$\hmpc~scale (upper
right) and a $3$\hmpc~scale (lower right).  Bars over the grayscale
maps indicate the local direction of the axis of structure and their
length is proportional to $\lambda_1'$.  Alignment of the axis of
structure with the $\lambda_1'$ minima suggests that the structures are
real on both scales (see Fig.~\ref{fig:GRFCompare2d_6}).  The
filament- or wall-like structures seen on a $10$\hmpc~scale are
divided into smaller filaments or walls on the $3$\hmpc~scale.
\label{fig:ScaleDemo}}
\end{figure}

We demonstrate the problem with this last approach using a
$15$\hmpc~deep slice from the SDSS Mr$20$ subsample
(Fig.~\ref{fig:ScaleDemo}).  At smoothing scales of both
$10$\hmpc~(upper right panel) and $3$\hmpc~(lower right panel), the
axis of structure appears to be aligned with the $\lambda_1'$ ridges,
suggesting that matter is condensed into filaments/walls on both
scales (see Fig.~\ref{fig:GRFCompare2d_6}).  Furthermore, the
filaments/walls found on the smaller scale appear to be components of
those on larger scales.  If the structures found on the
$10$\hmpc~scale were removed before any of the smaller structures were
identified, information would be lost.  Conversely, if the smaller
structures were all identified and removed first, there would be
little left to find the larger ones.

In this paper, we have presented results on a number of smoothing
scales and demonstrated how structure changed across those scales.
Filaments appear to be present on scales ranging from at least
$5$\hmpc~to $25$\hmpc~and are the dominant structures throughout most
of that range.  In Paper II, we extend these ideas by presenting a
method to identify and characterize individual filaments.  

\subsection{Walls, walls, everywhere}
\label{subsubsec:Walls}

There is evidence for a filament-to-clump progression as one smooths
on progressively smaller scales in the $z=0$ $\lambda$-space
projections in both simulations and real data.  Since smaller-scale
overdensities collapse before larger ones, they should be in later
stages of growth.  The collapse of a large-scale overdensity might be
expected to follow such a progression even if it were composed of
smaller-scale condensed objects (such as dark matter haloes).  The
coherence of these large-scale filaments and walls would presumably
depend upon the degree of fragmentation on intermediate scales.

\begin{figure}[t]
\epsscale{1.0}
\plotone{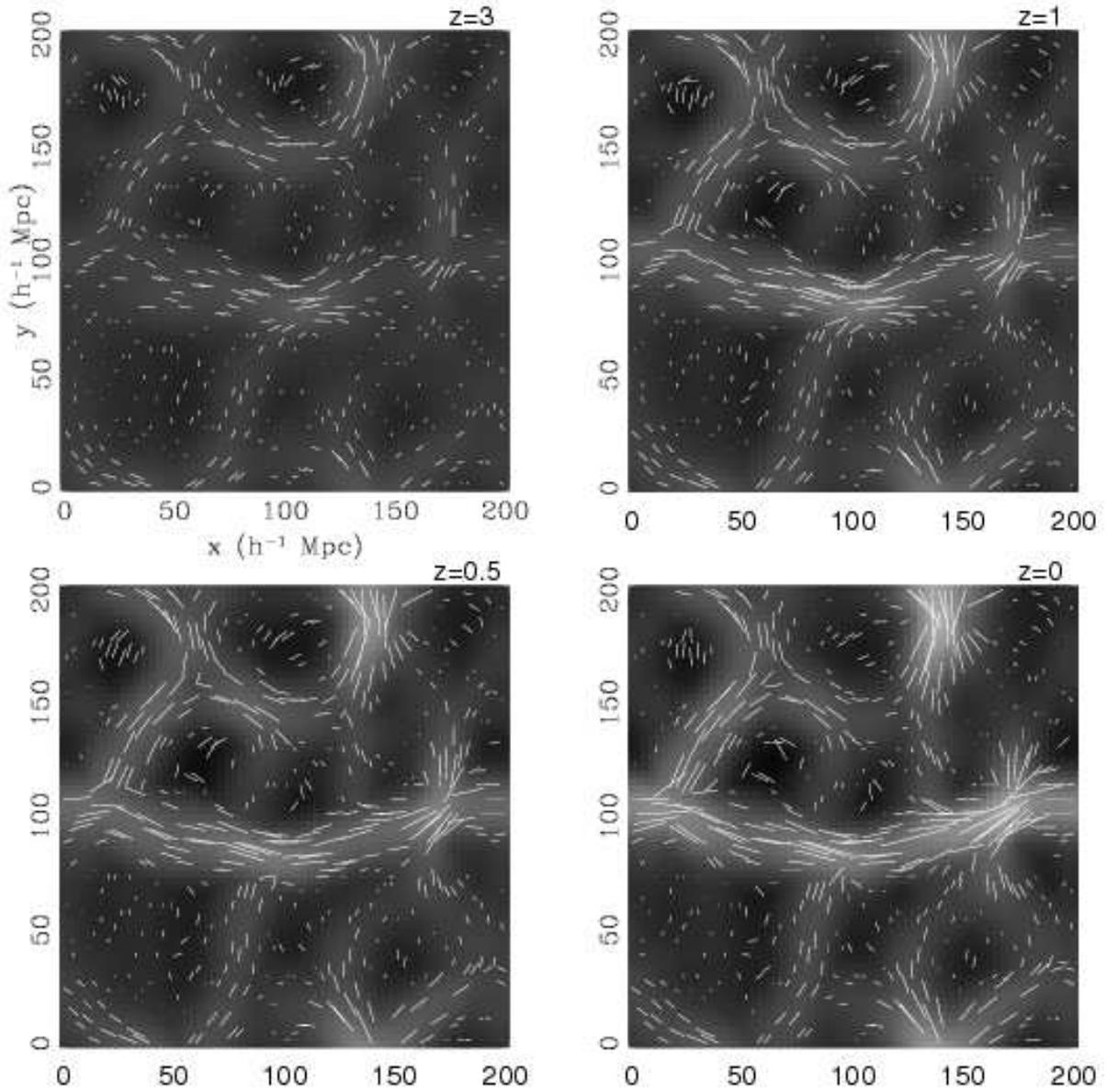}
\caption{The $\lambda_1'$ grayscale maps for a series of $10$\hmpc~deep slices from the dark matter distribution, very similar to those
in Fig.~\ref{fig:GRFCompare2d_6}.  The smoothing scale is
$15$\hmpc~and maps are plotted for $z=0$, $0.5$, $1$, and $3$.  Bars
indicate the direction of the axis of structure at random points on
the grid, and the bar length is proportional to the local value of
$\lambda_1'$.  In all maps shown here, the grayscale is scaled to the
minimum and maximum of $\lambda_1'$ in the $z=0$ map.  The axis of
structure aligns with minima in $\lambda_1'$ out to $z=3$,
unlike in gaussian random fields.
\label{fig:SimEvolve2}}
\end{figure}

One means of testing this picture is to look for walls in the very
early stages of non-linear evolution.  The $z=0$ $\lambda$-space
projections of the dark matter distribution show no convincing
evidence for wall-like structures at scales as large as $15$\hmpc,
but they may still exist on larger scales or earlier times.  The
former is difficult to explore in our $200$\hmpc~simulation box, but
we do have simulation outputs out to $z=3$.  In
Fig.~\ref{fig:SimEvolve2}, $\lambda_1'$ maps for a
$15$\hmpc~smoothing length are plotted for a range of redshifts, along
with bars indicating the axis of structure.  Even at $z=3$, it is
clear that, unlike in a gaussian random field, the axis of structure
aligns with the $\lambda_1'$ edges (though it is impossible to
distinguish filaments from walls in a two-dimensional projection).
As discussed in \S~\ref{sec:DSpace} and \S~\ref{subsubsec:Scale},
there appears to be a progression in both time and smoothing scale of
wall-to-filament-to-clump.  If this progression holds at $z=0$, we
might expect to see wall-like structures dominating on $\sim
40$--$50$\hmpc~scales.

\begin{figure}[t]
\plotone{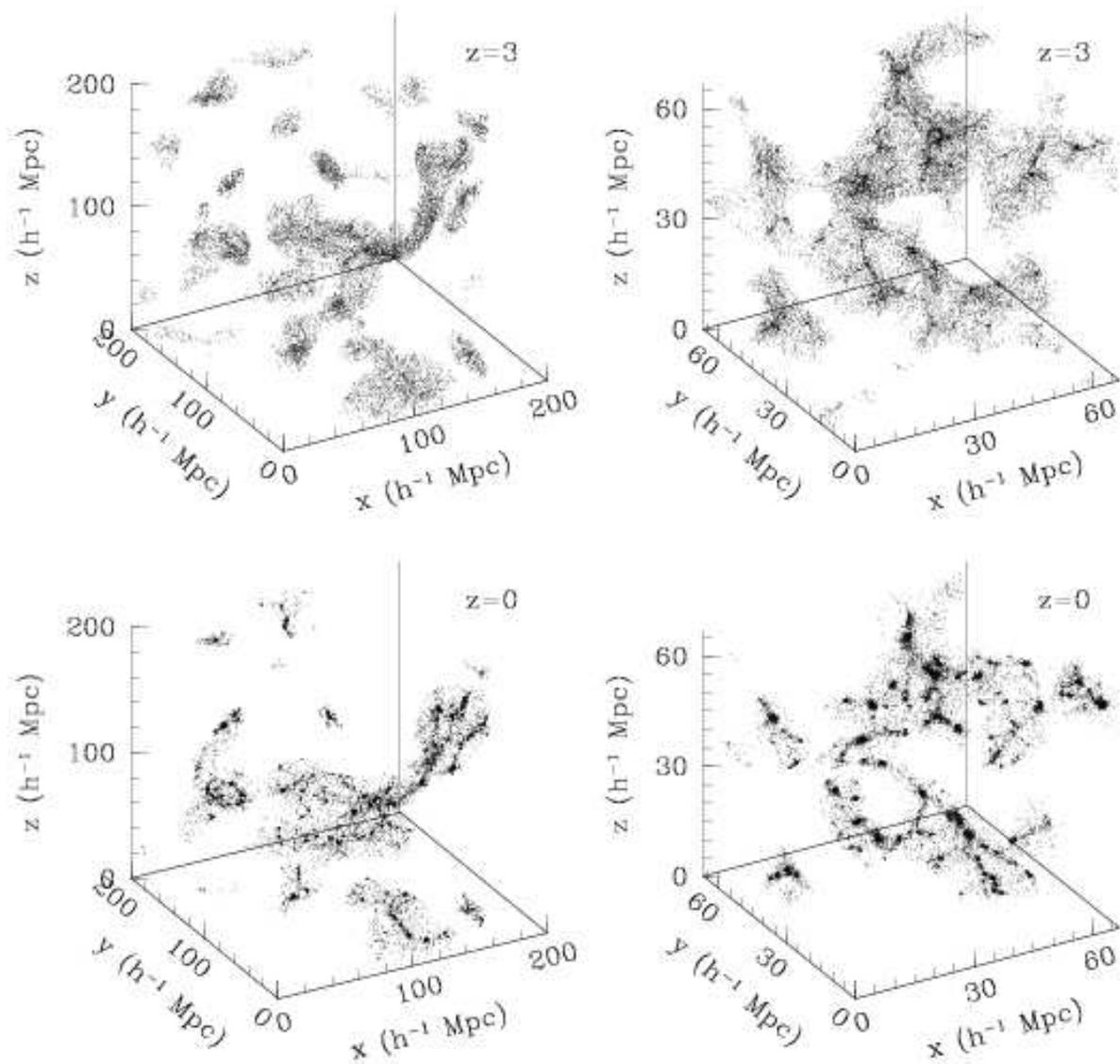}
\caption{
Selected dark matter particles in a $200$\hmpc~simulation box at
$z=0$ and $z=3$.  In the left column, particles are selected by the local value of
$\lambda_1'/\sigma_{15}(z)$ after $15$\hmpc~gaussian smoothing.  This
cut follows individual structures as they evolve.  Most of
the structures are sheets/walls at $z=3$, but have collapsed into filaments
by $z=0$.  In the right column, $5$\hmpc~smoothing is used in a $67$\hmpc~box.
On that scale, most of the structures are filamentary at $z=3$, but have collapsed into
clumps by $z=0$.\label{fig:Collapse}}
\end{figure}

The panels in the left-hand column of Fig.~\ref{fig:Collapse} show a subsample of dark
matter particles in the $200$\hmpc~simulation box satisfying
$\lambda_1'<l_{\rm cut}\sigma_{15}(z)$, where $\sigma^2_{15}(z)$ is the
variance of the density field at a given redshift when smoothed on a
$15$\hmpc~scale.  Here, $\lambda_1'$ is measured with
$15$\hmpc~smoothing and $l_{\rm cut}$ is a dimensionless constant.  The
cut isolates very similar regions of space at each redshift, but the
distribution of matter within those structures evolves a great deal
with time.  The regions isolated by the cut are very sheet-like, and
at $z=3$ the matter is smoothly distributed within the structures.  As
the evolution progresses, they collapse into filamentary and
clump-like structures.

Similarly, the right-hand column Fig.~\ref{fig:Collapse} shows
particles in regions of small $\lambda_1'$, but in a smaller box and
on a smoothing scale of $5$\hmpc.  Here, structures have begun to
collapse into filaments at $z=3$ and by $z=0$ are primarily
concentrated in clumps.  Despite the evolution of the structures
within the regions selected by a cut in $\lambda_1'$, the sizes and
morphologies of the regions themselves change very little -- they are
still pancake-like at $z=0$.  This suggests that, even at low
redshift, there should be evidence for wall-like structures in the
galaxy and dark matter distributions on large scales, even if these
walls are themselves made up of prominent filaments and clumps.  As
such, at all redshifts probed here, it can be the case that the
majority of the mass in the universe is in walls on large scales, which are
made up of filaments on smaller scales, which are made up of
clumps on still smaller scales.

\subsection{Conclusions}
\label{subsec:Results}

With the eigenvalues and eigenvectors of the Hessian matrix, we can
determine the type and orientation of structures in a continuous two-
or three-dimensional density field.  The `fingerprints' of clumps,
walls, and filaments in a three-dimensional space of eigenvalues
($\lambda$-space) provide templates for comparison to the
large-scale matter distribution.  In addition, as a result of
non-linear growth of structre, the third Hessian eigenvector (which
is used to define the `axis of structure') aligns with the axis of
filaments in both two and three dimensions, retaining this alignment
within approximately one smoothing length from the central axis.

We drew three volume-limited subsamples from the northern portion of
the SDSS spectroscopic survey (using the NYU-VAGC catalogue) and
computed their Hessian parameter distributions on a range of length
scales.  These distributions were then directly compared to those
found in a series of redshift-space mock galaxy catalogues generated
from a cosmological simulation using the concordance cosmology.
Results from this analysis include, 
\begin{itemize}
\item{ The SDSS galaxy distribution is different from a gaussian random
field on all smoothing scales used here (up to $25$\hmpc), showing
evidence for some combination of clusters, filaments, and walls on all
of these scales.  }
\item{ Filaments are the dominant structures on smoothing scales of $\sim
10$\hmpc~to $\sim 25$\hmpc, though there is some evidence for a
transition to wall-dominance at $\sim 25$\hmpc.  At $5$\hmpc, clumps
begin to dominate for the highest-contrast structures, but filaments are still apparent in the $\lambda$-space distributions.}
\item{In very negatively curved regions of the galaxy density field (as would be found near walls, filaments, and clusters), the direction of least curvature, indicated by the Hessian eigenvectors, points to other very negatively curved regions.  In other words, the axis of structure aligns with minima in $\lambda_1'$.  This occurs in projected slices at all smoothing scales studied here, providing a signature of non-linear growth even when non-linearities are not apparent to the eye in the density field.}
\item{ The $\lambda$-space projections of the SDSS galaxy distribution are
very similar to those expected in a $\Lambda CDM$ universe with gaussian random phase initial conditions.  There is some evidence that the data
are clumpier than the simulations.}
\item{ In projected slices, the $\lambda_1'$ maps of the galaxy distribution
are morphologically very similar to those of a gaussian random field.
Since filaments and walls are oriented along minima in
$\lambda_1'$, this suggests that the outline for the filament network
can be determined from the initial conditions.  }
\end{itemize}

To complement the results presented above for the SDSS galaxy
distribution, we generated $\lambda$ parameters at six redshifts in a
$\Lambda CDM$ cosmological N-body simulation, tracing the evolution
of the matter distribution from $z=3$ to $z=0$.  On comoving smoothing
scales of $5$\hmpc, filament-dominance at $z=3$ gives way to
clump-dominance at $z=0$.  At $z=3$ and on comoving scales of
$15$\hmpc, the dark matter distribution is dominated by low-contrast
walls, despite being only barely distinguishable from a gaussian
random field in $\lambda$-space.  Structure on this smoothing scale
becomes filament-dominated by $z=0$.  Finally, in projected slices,
the axis of structure follows the minima in $\lambda_1'$ on
smoothing scales as large as $15$\hmpc~and epochs as early as $z=3$.
The same alignment is {\it not} seen in gaussian random fields.

The $\lambda$-space distributions of the galaxy density field are
only a first step towards a more complete description of large-scale
structure.  In Paper II \citep{Thesis2}, we describe a method for
finding individual filamentary structures on a given scale, and then
quantitatively evaluate the connection between filaments and galaxy
clusters.  Eventually, we hope to both extend our morphological
studies to larger redshifts and to characterize the galaxy populations
in structures on a range of scales.  We will compare the SDSS
$\lambda$-space distributions to those found in hydrodynamic
simulations, test the dependence of the $\lambda$-space distributions
on the cosmological model and galaxy HOD, look for a dependence of
galaxy properties on the local geometry \citep{Park07}, and compare
the galaxy $\lambda$-space distribution to that of the hot gas
\citep[e.g.][]{ArielleThesis}.  

\section{Acknowledgments}
Funding for the SDSS and SDSS-II has been provided by the Alfred
P. Sloan Foundation, the Participating Institutions, the National
Science Foundation, the U.S. Department of Energy, the National
Aeronautics and Space Administration, the Japanese Monbukagakusho, the
Max Planck Society, and the Higher Education Funding Council for
England. The SDSS Web Site is http://www.sdss.org/.

The SDSS is managed by the Astrophysical Research Consortium for the
Participating Institutions. The Participating Institutions are the
American Museum of Natural History, Astrophysical Institute Potsdam,
University of Basel, University of Cambridge, Case Western Reserve
University, University of Chicago, Drexel University, Fermilab, the
Institute for Advanced Study, the Japan Participation Group, Johns
Hopkins University, the Joint Institute for Nuclear Astrophysics, the
Kavli Institute for Particle Astrophysics and Cosmology, the Korean
Scientist Group, the Chinese Academy of Sciences (LAMOST), Los Alamos
National Laboratory, the Max-Planck-Institute for Astronomy (MPIA),
the Max-Planck-Institute for Astrophysics (MPA), New Mexico State
University, Ohio State University, University of Pittsburgh,
University of Portsmouth, Princeton University, the United States
Naval Observatory, and the University of Washington.

We thank Jerry Ostriker for his many helpful comments and suggestions,
Michael Blanton for his hard work and help with VAGC, and Jim Gunn,
Neta Bahcall, and J. Richard Gott III for serving on the committee to the
thesis of which this work was a part.

{}
\newpage
\begingroup
\begin{deluxetable}{ccccc}
\tabletypesize{\footnotesize}
\tablewidth{0pc}
\tablecolumns{5}
\tablecaption{SDSS Subsamples}
\tablehead
{ 
\colhead{Sample} &
\colhead{Limiting $M_r$} &
\colhead{Dimensions} &
\colhead{$N_g$} &
\colhead{Smoothing Scales} \\
&
&
\colhead{($h^{-1}$~Mpc)} &
&
\colhead{($h^{-1}$~Mpc)} \\
\colhead{(1)} &
\colhead{(2)} &
\colhead{(3)} &
\colhead{(4)} &
\colhead{(5)} 
}
\startdata
Mr20 & $-20$ & $80\times 80\times 220$ & $7105$ & $5$, $7$, $10$ \\
Mr205 & $-20.5$ &  $140\times 140\times 340$ & $14048$ & $5$, $10$, $15$ \\
Mr21 &  $-21$ & $170\times 170\times 400$ & $11499$ & $15$, $20$, $25$ \\
\enddata
\label{tab:DataSamples}
\end{deluxetable}
\endgroup
\clearpage
\clearpage
\clearpage
\clearpage
\clearpage
\clearpage
\clearpage
\clearpage
\clearpage
\end{document}